\documentclass{emulateapj}
\newcommand{\myemail}{courteau@astro.queensu.ca}
\newcommand{\etal}{et~al.~}

\newcommand{\kms}{\ifmmode\,{\rm km}\,{\rm s}^{-1}\else km$\,$s$^{-1}$\fi}
\newcommand{\magarc}{\ifmmode {{{{\rm mag}~{\rm arcsec}}^{-2}}}
             \else {{{mag}$~${arcsec}$^{-2}$}}
             \fi}

\def\Equ#1{Eq.~(\ref{eq:#1})} 
\def\se#1{\S\ref{sec:#1}}
\def\Fig#1{Fig.~\ref{fig:#1}}

\def\be{\begin{equation}} 
\def\ee{\end{equation}} 
\def\ifm#1{\relax\ifmmode#1\else$\mathsurround=0pt #1$\fi}

\def \spose#1{\hbox to 0pt{#1\hss}}
\def \lta{\mathrel{\spose{\lower 3pt\hbox{$\sim$}}
     \raise 2.0pt\hbox{$<$}}}
\def \gta{\mathrel{\spose{\lower 3pt\hbox{$\sim$}}
     \raise 2.0pt\hbox{$>$}}}
\def\ltsima{$\; \buildrel < \over \sim \;$} 
\def\lsim{\lower.5ex\hbox{\ltsima}} 
\def\gtsima{$\; \buildrel > \over \sim \;$} 
\def\gsim{\lower.5ex\hbox{\gtsima}}

\def\prop{\propto}

\def\kms{\ifmmode\,{\rm km}\,{\rm s}^{-1}\else km$\,$s$^{-1}$\fi}
 
\def\kpc{\,{\rm kpc}} 
\def\Mpc{\,{\rm Mpc}} 

\def\Msun{M_{\odot}}
\def\lsol{L_{\odot}}

 \def\omb{\Omega_{\rm b}}

\def\rhoc{\rho_{\rm crit}}
\def\fbar{f_{\rm bar}} 

\def\Vv{V_{\rm vir}}
\def\Mv{M_{\rm vir}} 
\def\Rv{R_{\rm vir}} 
\def\Dv{\Delta_{\rm vir}}
\def\Jv{J_{\rm vir}}
\def\Mh{M_{\rm h}}

\def\egf{\epsilon_{\rm gf}} 
\def\esf{\epsilon_{\rm sf}}

\def\Jd{J_{\rm d}}  
\def\jd{j_{\rm d}}  
\def\md{m_{\rm d}} 
\def\ld{\lambda_{\rm d}}

\def\Vobs{V_{\rm obs}}

\def\Rd{R_{\rm d}} 
\def\Md{M_{\rm d}}
\def\Ie{I_{\rm eff}}
\def\Re{R_{\rm eff}}
\def\ReK{R_{e,K}}

\def\mus {\mu_0} 
\def\mue {\mu_{\rm eff}}

\def\Upstilde{\widetilde{\Upsilon}} 
\def\Upsilond{\Upsilon_{\rm d}}
\def\Upsilons{\Upsilon_{\rm *}}
\def \Rx {R_{\rm d}} 
\def \V22{V_{2.2}} 
\def \r22{R_{disk}} 
\def \BI {B-I} 
\def \VI {V-I}

\def \dlogV {\Delta\log{V}} 
\def \dlogR {\Delta\log{R}} 
\def \dlogL {\Delta\log{L}} 
\def \dlog {\Delta\log}
\def \partialvr {\dlogV \thinspace / \thinspace\dlogR} 
\def \partialVRL {\dlog V|L \thinspace / \thinspace\dlog R|L} 
\def \partialRLV {\dlog R|V \thinspace / \thinspace\dlog L|V}
\def \partialVLR {\dlog V|R \thinspace / \thinspace\dlog L|R} 
\def \CVL { {\cal C}_{\rm VL}}
\def \CRL { {\cal C}_{\rm RL}}
\def \CRV { {\cal C}_{\rm RV}}

\def\pmb#1{\setbox0=\hbox{#1}%
\kern-.025em\copy0\kern-\wd0 \kern.05em\copy0\kern-\wd0
\kern-.025em\raise.0433em\box0}

\def \ion#1#2{#1{\footnotesize{#2}}\relax} 
\def \ha {H$\alpha$\ } 
\def \hi {\ion{H}{I} }

\def \farcm{\hbox{$.\mkern-4mu^\prime$}} 
\def \farcs{\hbox{$.\!\!^{\prime\prime}$}}

\def \littlemm{\ifmmode{\scriptscriptstyle m }
     \else{\hbox{$\scriptscriptstyle m $ }}\fi} 
\def \topemm{\raise .9ex \hbox{\littlemm}} 
\def \magpoint{\hbox to 2pt{}\rlap{\hskip -.5ex
     \topemm}.\hbox to 2pt{}} \def \deg {$^\circ$}

\shorttitle{Scaling Relations of Spiral Galaxies}
\shortauthors{Courteau et~al.}

\slugcomment{ApJ 2007, in press.} 
\begin{document}

\title{Scaling Relations of Spiral Galaxies}

\author{St\'{e}phane Courteau}

\affil{Department of Physics, Engineering Physics and Astronomy, 
Queen's University, Kingston, Ontario, Canada} 
\email{\myemail}

\author{Aaron A. Dutton} \affil{
UCO/Lick Observatory and Department of Astronomy and Astrophysics, 
University of California, Santa Cruz, CA}

\author{Frank C. van den Bosch} \affil{Max-Planck-Institut f\"ur Astronomie,
K\"onigstuhl 17, Heidelberg, Germany} 

\author{Lauren A. MacArthur}
\affil{Dept. of Astronomy, California Institute of Technology, 
 Pasadena, CA}

\author{Avishai Dekel} \affil{Racah Institute of Physics, The Hebrew
University, Jerusalem, Israel} 

\author{Daniel H. McIntosh} \affil{University of Massachusetts,
Department of Astronomy, Amherst, MA}

\and

\author{Daniel A. Dale} \affil{Department of Physics \& Astronomy,
University of Wyoming, Laramie, WY} 

\bigskip

\begin{abstract}
  We construct a large data set of global structural parameters for
  1300 field and cluster spiral galaxies and explore the joint
  distribution of luminosity $L$, optical rotation velocity $V$, and 
  disk size $R$ at $I$- and 2MASS $K$-bands.  We focus our
  study on scaling relations and residual correlations of benefit to
  galaxy structure and formation studies, rather than deriving
  precise distance estimators.
  The $I$- and $K$-band velocity-luminosity ($VL$) relations have 
  log-slopes of $0.29$ and
  $0.27$, respectively with $\sigma_{\rm ln}(VL) \simeq 0.13$, and
  show a small dependence on color and morphological type in the sense
  that redder, early-type disk galaxies rotate faster than bluer, later
  type disk galaxies for most luminosities.  The $VL$ relation at $I$- 
  and $K$-bands is independent of surface brightness, size and light
  concentration.
  The modeling of the size-luminosity ($RL$) relation is somewhat 
  sensitive to the adopted
  fitting method, due in part to strong dependences on surface
  brightness and color and significant uncertainties in the
  measurement of disk scale lengths.  The log-slope of the $I$- and
  $K$-band $RL$ relations is a strong function of morphology and varies
  from 0.25 to 0.5 with a mean of $0.32$ for all Hubble types.  At
  most luminosities, early-type disk galaxies have shorter scale lengths
  than later-type ones - this latter trend may be reversed at high
  luminosities.  The average dispersion $\sigma_{\rm ln}(RL)$
  decreases from 0.33 at $I$-band to 0.29 at $K$, likely due to the
  2MASS selection bias against lower surface brightness galaxies.
  Measurement uncertainties are $\sigma_{\rm ln V} \simeq 0.09$,
  $\sigma_{\rm ln L} \simeq 0.14$ and somewhat larger and harder to
  estimate for $R$.
  We show that the color dependence of the $VL$ relation is consistent
  with expectations from simple stellar population synthesis models; 
  the $RL$ relation shows a weak but opposite trend expected
  from those models, suggesting that at a given 
  stellar mass, smaller galaxies are redder.
  The $VL$ and $RL$ residuals are largely uncorrelated with each other
  with a correlation coefficient $r=-0.16$ and $\partialVRL =-0.07 \pm
  0.01$; the $RV-RL$ residuals show a weak positive correlation with 
  $r=0.53$. 
  These correlations suggest that scatter in luminosity is not 
  a significant source of the scatter in the $VL$ and $RL$ relations.
  The observed scaling relation slopes, zero-points, and scatters can
  be understood in the context of a model of disk galaxies embedded 
  in dark matter halos that invokes low mean spin parameters and dark
  halo expansion, as we describe in our companion paper (Dutton \etal
  2007).  We discuss in two appendices various pitfalls of standard
  analytical derivations of galaxy scaling relations, including the 
  Tully-Fisher relation with different slopes.  Our galaxy data base
  is available at www.astro.queensu.ca/$\sim$courteau/data/VRL2007.dat. 
\end{abstract}


\section{Introduction}\label{sec:intro}

Understanding the origin  and nature of galaxy scaling  relations is a
fundamental quest  of any successful theory of  galaxy formation.  The
success  of a  particular  theory will  be  judged by  its ability  to
predict  the  slope, scatter,  and  zero-point  of  any robust  galaxy
scaling relation at any  particular wavelength.  Some observed scaling
relations  in spiral galaxies,  based on  their size,  luminosity, and
rotation speed,  can be reproduced  {\it individually} to  fairly good
accuracy  by  invoking galaxy  formation  models  that include  virial
equilibrium after dissipational collapse of spherical cold dark matter
(CDM) halos and angular momentum  conservation (e.g. Mo, Mao, \& White
1998, hereafter MMW98; van den Bosch 1998, 2000, hereafter collectively
as vdB00; Navarro \& Steinmetz 2000, hereafter  NS00;
Firmani \& Avila-Reese 2000, hereafter FAR00; Shen, Mo, Shu 2002).

One of the most firmly established empirical scaling relations of disk
galaxies is the  Tully-Fisher relation (TFR; Tully \&  Fisher 1977); a
tight correlation between the  total luminosity and the rotation speed
of a disk galaxy.  We here refer to the TFR as the velocity-luminosity
($VL$) relation.  To date,  no single CDM-based model of galaxy
formation  can {\it simultaneously}  reproduce the  slope, zero-point,
scatter,  and color trends  of the  $VL$ relation,  match the  shape and
normalization  of the local luminosity function,  and  explain the  sizes,
colors, and metallicity of disk galaxies (see, e.g., vdB00; Bell~\etal
2003a;  Dutton  \etal 2007,  hereafter  D07).   In
addition, simultaneously accounting for  the mass and angular momentum
distribution of spiral  galaxies in a gas dynamical  context remains a
major challenge  for hierarchical  formation models (Navarro \& White
1994; van den Bosch \etal 2002b; Governato \etal 2007). A complete
theory of  galaxy scaling relations  awaits a fuller  understanding of
structure  forming mechanisms and  evolutionary processes  (e.g.  star
formation, merging,  feedback, and cooling  prescriptions) in galaxies. 
Likewise,  the  fine-tuning  of  these  galaxy
formation  and evolutionary  models demands  a careful  examination of
empirical scaling relations of galaxies.

In  order to  set  up a  framework  for the  study  of galaxy  scaling
relations, we  examine the correlations  of parameters related  by the
virial  theorem,  $V^2  \prop  M/R$.  We  consider  three  fundamental
observables  for  each disk  galaxy:  the  total  luminosity $L$,  the
stellar scale length  $\Rd$ of the exponential disk,  and the observed
circular velocity $V$.  The stellar mass, $M_*$, can be estimated from
the   luminosity   by   assuming   a  stellar   mass-to-light   ratio,
$\Upsilon_*=M_*/L$.  The size-luminosity ($RL$) relation of galaxy disks
is also expressed  as $L \propto \Sigma_\circ \Rd^2$,  where $\Sigma_\circ$ 
is the disk central surface brightness.

Key to  mapping fundamental dynamical  trends in spiral  galaxies, the
measurement of $VL$  and $RL$ relations and detection  of their correlated
residuals require  velocity amplitudes  measured at a  suitably chosen
radius representative  of the flat  part of resolved  rotation curves,
red/infrared magnitudes to minimize extinction and stellar population 
effects, accurate disk scale lengths, and, ideally, color terms from 
digital imaging with a  broad baseline (e.g. $B-K$) to test for $\Upsilon_*$
variations in  the stellar population and extinction  effects.  One of
the goals of this paper is to  assemble such a data base for field and
cluster galaxies (within the limits of available material).

A  study of  scaling relations  in  irregular and  spiral galaxies  by
Salpeter \&  Hoffman (1996) yielded  the correlations
$L_B  \propto  R^{2.68}  \propto  V^{3.73} \propto  M_{\rm  HI}^{1.35}
\propto M_{\rm  dyn}^{1.16}$, where  $M_{\rm HI}$, $M_{\rm  dyn}$, and
$R$, are  the \hi and  dynamical masses, and a  characteristic radius,
respectively.   The blue  luminosities,  as used  in  that study,  are
notoriously  sensitive  to  dust  extinction  and  stellar  population
plus dynamical effects cannot  be simply isolated.  A new study
of scaling relations in  the (near-)infrared would provide more robust
dynamical  constraints  to  galaxy  formation models.   While  the $VL$
relation has been examined  at nearly all optical-IR wavelengths (e.g.
Strauss \& Willick 1995;  Verheijen 2001, hereafter V01; Masters \etal
2006; Pizagno \etal 2007, hereafter P07), comparatively few analyses 
of the $RL$ relation of
spiral galaxies have been  reported so  far (Salpeter \&  Hoffman 1996;
de Jong \& Lacey 2000, hereafter dJL00; Shen \etal 2003, hereafter Sh03). 
This is partly because the accurate disk scale lengths needed to calibrate 
the $RL$ relation, at any wavelength,
have  only  recently  become   available  for  large  databases  (e.g.
Courteau  1996;  Dale  \etal  1999; MacArthur  \etal  2003,  hereafter
MCH03).  Furthermore,  most investigations of the $VL$  and $RL$ relations
have  used grossly  incomplete databases  owing to  the nature  of the
sample (e.g.  single cluster)  or selection limits (magnitude, surface
brightness, diameter,  etc.)  The  availability of light  profiles for
very  large galaxy  databases (e.g. {\sl Sloan  Digital  Sky Survey},
York \etal 2000 [hereafter SDSS] and the {\sl Two Micron All-Sky Survey},
Skrutskie \etal 1997 [hereafter 2MASS]) heralds a  new era  for the
study of galaxy scaling relations (e.g. Shen \etal 2003; P07) with
tractable selection biases.  

The empirical $VL$ relation is expressed as
\be L \propto V^a , 
\ee
with the near-IR log-slope $a  \simeq3.0 \pm 0.2$ (Willick \etal 1997;
Giovanelli \etal  1997, hereafter G97; Courteau 1997; hereafter C97;
Courteau \etal 2000; V01; Masters \etal 2006).
Reported values of the log-slope $a$ range from 2.8 in the blue to 4.0
in the infrared  (Willick \etal 1997; Tully \&  Pierce 2000; V01; P07)
for both high  and low surface brightness galaxies  (Zwaan \etal 1995;
V01).  However, log-slopes  steeper than $a \sim 3.5$  in the infrared
typically result from small samples and excessive pruning on the basis
of idealized morphology or kinematics, a narrow range of inclinations,
redshift cutoffs, etc. (Bernstein  \etal  1994;  V01;  Kannappan,
Fabricant, \&  Franx 2002, hereafter  KFF02). The slope,  scatter, and
zero-point of blue $VL$ relations are predominantly dominated by stellar
population and  dust extinction effects (e.g. Aaronson \& Mould 1983;
G97;  Willick \etal 1997; Tully \& Pierce 2000; P07)  and on the
techniques  used to  recover  the major  observables and fitting for
fundamental relations (e.g. Strauss \& Willick 1995; C97; V01; 
Bell \& de Jong  2001; KFF02). Because we are 
mainly interested  in masses, rather than luminosities, we do not 
concern ourselves with the $VL$ and other scaling relations measured 
at blue wavelengths.

It is shown  in Appendix B that previous  dynamical derivations of the
observed $VL$  relation which  arrived at log  slopes of $3$  ($-7.5$ in
magnitudes)   based   on   the   virial   relation   $\Mv\propto\Vv^3$
(e.g., MMW98; NS00) and $4$ ($-10$ in magnitudes) based  on  disk
dynamics used erroneous assumptions.  A complete physical
interpretation  of the  $VL$ relation is still missing (for a fuller
discussion, see e.g., Gnedin \etal 2006; D07). In terms of
observables, the  most fundamental relation is between  the total {\it
baryonic} mass of a galaxy, inferred via its infrared luminosity and
a stellar mass-to-light  ratio and total gas mass (\hi  + He), and its
total  mass inferred  from  the asymptotic  circular  velocity of  the
galaxy disk (McGaugh \etal 2000; Bell  \& de Jong  2001; V01; McGaugh
2005; Geha \etal  2006; Gurovich  2006; De  Rijcke \etal  2007).  The
``baryonic'' $VL$ relation is expressed as
\be {\cal M}_{\rm{bar}} \propto V^{a_{\rm bar}} . 
\label{eq:barTF}
\ee

Since  disk  gas mass  fractions  typically  increase with  decreasing
luminosity (e.g., McGaugh \& de Blok 1997),  one typically  has 
$a_{\rm bar}  > a$.  The slope $a_{\rm bar}$ is sensitive  to the
method used to determine  stellar mass-to-light ratios.  Using stellar
populations models one finds $a_{\rm bar} \simeq 3.5$ (Bell \& de Jong
2001;  McGaugh  2005),  while  adopting stellar  mass-to-light  ratios
derived  from  MOND  results,  by  construction,  in  $a_{\rm  bar}=4$
(e.g. McGaugh 2005).

While the log-slope  of the $VL$ relation can  be reproduced fairly well
by naive  derivations from CDM-based structure  formation models (e.g.
MMW98; NS00; D07;  see also Appendix B), the  predicted scatter can be
large compared to the inferred  ``cosmic'' scatter of $\lsim 0.25$ mag
in   red/infrared   bands  (Willick   \etal   1996;   V01; P07)   and
interpretations about its dependence  differ (see below).  Besides the
basic  understanding of the  slopes of  galaxy scaling  relations, the
dependence  of  their  scatter   has  also  been  addressed  by  many,
especially for  the $VL$ relation  (e.g. Aaronson \& Mould  1983; Giraud
1986; Rhee 1996; Willick \etal 1997; KFF02; Courteau \etal 2003), and 
can be  used to set
realistic constraints  on structure formation models  (Courteau \& Rix
1999,  hereafter CR99; Heavens  \& Jimenez  1999;  FAR00; NS00;  V01;
Buchalter, Jimenez, \& Kamionkowski  2001; Shen,  Mo,  \& Shu  2002;
Gnedin \etal 2006;  D07).  While various trends in  the scatter of the
blue $VL$ relation have been  discussed in the past, correlations of the
near-infrared $VL$ residuals  with inclination, size, concentration, gas
fraction,  or far infrared  luminosity are  only few  and inconclusive
(Aaronson \& Mould  1983; V01; P07). The dependence  of the scatter in
the near-IR relation on color and surface brightness is, however, still
a matter of contention that we discuss in \se{color}.  The scatter in
the $RL$ relation is also addressed in \se{color}.
 
The study of scaling relations in galaxies has benefited from the
two-pronged application of the $VL$ relation for the purposes of: (i)
Estimating relative distances to measure deviations from the mean
Hubble flow (see, e.g., Strauss \& Willick 1995 and the reviews in
Courteau, Strauss, \& Willick 1999); and (ii) testing galaxy formation
and evolution models (Dalcanton, Spergel \& Summers 1997; MMW98;
vdB00; D07).  The philosophy of sample selection and calibration
differs in both cases. For cosmic flow analyses, the calibration and
science samples must be pruned mainly on the basis of morphology in
order to minimize systematic errors and ensure the smallest possible
magnitude (distance) error.  Since scatter in the $VL$ relation depends
strongly on the slope of the relation, it is found that the
combination of steepness, magnitude errors, extinction correction, and
sky stability favors red ($R \& I$) bands for smallest distance errors
and cosmic flows applications (C97; Giovanelli \etal 1999; V01).  
The accuracy of bulk flow solutions also
depends on the size of the sample.  In order to collect large enough
samples, $VL$ calibrations for flow studies rely mostly on \hi line
widths or \ha rotation curves that can be collected relatively quickly
on modest aperture telescopes.  These rotation measures based on radio
and optical spectra typically sample the disk rotation out to 
${\sim}3-4$ disk scale lengths and are readily calibrated on the same 
system (C97; Dale \etal 1999).
 
In contrast, the use of the $VL$ relation as a test bed for galaxy
formation models requires the widest range of morphological types, to
sample all structural properties, and that extinction and stellar
population effects be tractable to isolate genuine dynamical
correlations.  The nearly dust-insensitive infrared bands are thus
ideal for such applications (V01).  Rotation velocities are preferably
extracted from fully resolved \hi rotation curves obtained using
aperture synthesis maps that sample the disk rotation out to roughly
$4-5$ disk scale lengths.
 
Ideally, the  study of scaling  relations should rely on homogeneous
complete volume-limited samples assembled with the very purpose of
testing  for  broad  structural  and dynamical  differences;  at  the
moment,  we must contend  with the  more finely  pruned heterogeneous
samples of early and mostly late-type spirals that have been collected 
during the last decade mostly for flow studies. It is currently not  
possible to correct any of the existing catalogs for incompleteness 
or assemble a volume-limited sample extending to low mass that is 
complete in any meaningful sense (e.g. McGaugh  2005).  The latest  
heterogeneous data bases, which may include near-infrared luminosities  
and colors, for large samples of spiral galaxies still enable us to 
characterize the dependence of scaling relations under the assumption 
that the different selection biases even out when sufficiently 
many different samples are collected together, and to examine various 
constraints of structure formation models in ways hitherto unsettled.
 
Courteau \etal (2003) showed that  barred and unbarred  galaxies have
similar dynamical properties (beyond  the co-rotation radius) and that
they share the same $VL$ relation.  As a natural extension of this study
and  CR99, we  use in \se{virial} and \se{color} extensive  all-sky 
distance-redshift catalogs presented in \se{data} to characterize the 
mean $VRL$ disk galaxy relations. 
In an attempt to extend our analysis
to  the  (nearly)  dust-free  domain,  we consider  in  \se{IRTF}  the
infrared photometry from  2MASS for a large sample  of spiral galaxies
and discuss any relevant caveats.
Although the  2MASS misses low  luminosity and low  surface brightness
galaxies it  still serves  some purposes for  our analysis  of scaling
relations.
In \se{scatter} we bolster the notion of surface brightness and size
independence of the $VL$ relation, and we address in \se{CR07} the scatter 
of the $VL$ and $RL$ relations and their weak dependence on disk color.  
The main results are summarized in \se{discussion}. Four appendices 
give A) figures for the mean $VRL$ 
relations for each data sample, B) a derivation on the origin 
of disk galaxy scaling relations, C) an alternative derivation of 
the Tully-Fisher relation, and finally D) color transformations. 

The collected data and basic observations in this paper form the basis
of the analytical modeling of spiral galaxy scaling relations in our
companion paper (D07.  Our basic data base is available publicly 
at www.astro.queensu.ca/$\sim$courteau/data/VRL2007.dat. 

\section{Available Data and Basic Analysis}{\label{sec:data}}

We consider four major samples for which accurate near-infrared galaxy
observables and rotational velocities are available.  These include
(a) the large $I$-band survey of galaxy distances for bright field
spirals in the Southern sky by Mathewson, Ford, \& Buchhorn (1992;
hereafter ``{\bf MAT}''); the all-sky $I$-band $VL$ surveys of cluster
and field spirals by (b) Dale \etal (1999; hereafter ``{\bf SCII}''),
(c) Courteau \etal~(2000; hereafter ``{\bf Shellflow}''); and (d),
the multi-band $BRIK$ $VL$ survey of Ursa Major
cluster galaxies by Tully \etal (1996) and V01 (hereafter ``{\bf
UMa}'').  The first three $VL$ surveys were originally designed to map
the convergence of the velocity field within ${\sim}60h^{-1}$ Mpc, with
the SCII and Shellflow studies paying special attention to VL
calibration errors for data from different observatories.  By design, these
surveys favor bright late-type galaxies with inclinations on the sky
greater than 45\deg; most have $i\simeq 60$\deg.
 
Properties  for each sample  are given  in Table  1. These  include in
Col.~(2), the number of galaxies from the original database that have
the full complement of relevant observables; Col.~(3), the  nature of the
sampled galaxies (cluster  or  field;  the  predominantly  ``field''
MAT and Shellflow surveys include  a small  fraction of  cluster
galaxies); Col.~(4), the original photometric band coverage.  $B$-band
magnitudes for  all the MAT galaxies  were extracted from  the RC3 (de
Vaucouleurs \etal 1991).   We also have SDSS $g-i$  colors for 39 SCII
galaxies (see Appendix C) and $JHK$ 2MASS Kron magnitudes and colors for
the brightest $360$ SCII galaxies; Col.~(5), the magnitude or diameter
limits of the  original catalog; Col.~(6), the redshift  limits of the
survey in \kms.  For UMa, we  adopted the cluster distance of 20.7 Mpc
($HST$ Key Project,  Sakai \etal 2000); and in  Col.~(7), the rotation
measure, extracted from spatially resolved \hi or \ha rotation curves.
 
While  the  MAT  and  SCII  samples  used  both  \hi  linewidths  from
integrated spectra and \ha  rotation velocities from resolved rotation
curves,  we  use  only  the  sub-samples of  galaxies  with  available
H$\alpha$, which  is largest  for both samples.   For the  purposes of
galaxy  structure studies H$\alpha$  rotation velocities  are prefered
over \hi  linewidths as the rotation  velocity is measured  at a known
point of the rotation  curve.  Furthermore, \hi linewidths suffer from
uncertain turbluence corrections.

Rotation  velocities  for  the  Shellflow sample  were  measured  from
resolved \ha rotation curves.  For the UMa database, we use the sample
of 38 galaxies with $V_{\rm  max}$ measured from resolved \hi rotation
curves (V01).

Disk  scale length  measurements were  computed in  somewhat different
ways  for  each  sample.  For  the  MAT  galaxies,  we have  used  the
two-dimensional bulge-disk (B/D) decompositions of MAT $I$-band images
by Byun (1992).  These decompositions assumed a {de~Vaucouleurs} bulge
profile  which  is  not  ideal  for  late-type  galaxies  (MCH03)  and
therefore the  extrapolated disk  central surface brightnesses  may be
biased low  by a few tenths  of a magnitude.  Those  scale lengths are
however deemed adequate for the  purpose of our study since they match
the mean $RL$ relation (see  \se{virial}).  SCII disk scale lengths 
were obtained by fitting a straight line to the exponential part of the
$I$-band  surface brightness  profile from  ${\sim}21$ to
${\sim}25$ $I$-\magarc (the so-called ``marking the disk'' technique).
Shellflow  scale  lengths  were  extracted  from  one-dimensional  B/D
decompositions  of  azimuthally-averaged  $I$-band surface  brightness
profiles  (Courteau \etal  2000;  MCH03).  These  fits  account for  a
S\'ersic bulge  and an exponential  disk.  Disk scale lengths  for the
UMa sample were measured from a ``marking the disk'' technique but the
fit baseline  is unspecified  and erratic fits  are reported  in Tully
\etal (1996).  The latter could be  due to inclusion of bulge light in
the disk fit (scale lengths  biased low) for the brighter galaxies and
sky  underestimate  (scale  lengths   biased  high)  for  the  fainter
galaxies.  However, much like MAT, these peculiarities  do not affect
the  UMa $RL$  relations  as  compared against  the  other samples 
(see \se{virial} and Appendix A).

The conversion of apparent observables to physical parameters uses 
the distance scale adopted by the original authors, except for UMa 
galaxies for which we adopted $D=20.7$ Mpc (Sakai \etal 2000). 

Deviations  from  a smooth  Hubble  flow  may  cause distance  errors,
especially  for the  nearby galaxies  in  UMa.  For  the more  distant
galaxies in  MAT, Shellflow, and  SCII, typical thermal  velocities of
${\sim}200$ \kms\  would cause relative distance error of ${\sim}3\%$.
We find  in \se{corrections} that distance errors  which affect mainly
the luminosities and sizes are  small compared to the other sources of
observational  uncertainty.
For more details about each sample, 
the reader is referred to the original papers. 

\subsection{Corrections}{\label{sec:corrections}}

In order to homogenize the four data samples, we have applied 
the inclination corrections for the line widths, luminosities, 
and scale lengths in a uniform way, as described below.

The rotation velocity estimates are corrected for inclination and 
redshift broadening using

\be
 V = V_{\rm obs} / {[(\sin i) (1+z)]}. 
\ee

The inclination angle is derived via

\be
 i = \cos^{-1} \sqrt{{(b/a)^2 - q^2_0}\over{1-q^2_0}},  
\ee

where  the semi-minor  and  semi-major axes  are  determined from  CCD
isophotal  fitting, and  $q_0$ is  the intrinsic  flattening  ratio of
edge-on spiral  galaxies, here taken  to be $q_0=0.2$ (e.g.  Haynes \&
Giovanelli 1984; Courteau 1996; Tully \etal 1998; 
Sakai \etal 2000).

The  apparent  ($I$-band)  magnitudes,  $m_I$, are  corrected  for  both
internal  extinction,  $A_{\rm int}$,  and  external (i.e.,  Galactic)
extinction, $A_{\rm ext}$. A  small k-correction, $A_{\rm k}$, is also
applied such that
\begin{equation}
m_I = m_{I,{\rm obs}} - A_{\rm int} - A_{\rm ext} - A_{\rm k}.
\end{equation}
Internal  extinctions  are  computed  using the  line  width  ($W=2V$)
dependent relation from Tully \etal (1998):
\begin{equation}
A_{\rm  int}   =  \gamma_I(W) \log(a/b)  
               =  \left[0.92 + 1.63(\log  W -2.5)\right] \log(a/b)
\end{equation}
where $a/b$ is the major-to-minor axis ratio and $\gamma_I(W)$ is 
a line width dependent extinction parameter.  The external  (Galactic)
extinction  is computed  using the  dust extinction  maps  of Schlegel
\etal  (1998), while  the k-corrections  are computed  using  the line
width dependent formalism of Willick \etal (1997).

The absolute magnitudes, $M$, are computed using
\begin{equation}
  M = m - 5 \log D_{\rm L} -25; \;\;\;\; D_{\rm L}=\frac{V_{\rm
      CMB}}{100h}(1+z),
\end{equation}
with  $V_{\rm  CMB}$  the  systemic  velocity of  the  galaxy  in  the
reference frame  at rest with  the cosmic microwave  background (Kogut
\etal 1993).   The $I$- and  $K$-band luminosities are  computed using
for the Sun $M_{I,\odot}=4.19$ and $M_{K,\odot}=3.33$.

Disk scale lengths from $I$-band images are corrected for inclination using
\begin{equation}
R_{\rm I} = R_{\rm I,obs} / [1+0.4\log(a/b)],
\label{eq:Giov}
\end{equation}
(Giovanelli  \etal 1994),  and converted  into kpc  using  the distance
$D_{\rm L}$.  The correction above (\Equ{Giov}) is  somewhat uncertain 
but the choice of observed or deprojected scale lengths does not affect
our final results.

Finally, central  surface brightnesses are  corrected for inclination,
Galactic  extinction,  and cosmological  dimming  (per unit  frequency
interval) using
\begin{equation}
\mu_{0,I} = \mu_{0,I,\rm obs} + 0.5 \log(a/b) - A_{\rm ext} - 2.5\log(1+z)^3.
\end{equation}
The coefficient $0.5$  in front of the extinction  term $\log(a/b)$ is
empirically determined by demanding that the residuals of the relation
between the $I$-band central  surface brightness and rotation velocity
has no  inclination dependence\footnote{For a disk of zero thickness,
one expects this factor to range from 0 for an optically thick disk
to 2.5 for an optically thin disk.}.  We  have  assumed  full
transparency in the $K$-band.  Following Giovanelli \etal  (1997) we
assume an uncertainty of 15\% in $\gamma_I$, $a/b$, $A_{\rm ext}$, and
$A_{\rm   k}$,   and  propagate the errors. Including distance
uncertainties, the average errors  on the observables are $\sigma_{\ln
  V}  \simeq 0.09$,  $\sigma_{\ln  L} \simeq  0.14$, and  $\sigma_{\ln
  R}=0.15$  (adopting a  fitting  uncertainty of  15\%  on disk  scale
length).  

Distance errors in the $VL$ relation distance affect mainly luminosity.  
If we include a distance uncertainty in terms of the recession velocity,
the TF scatter is dominated by the uncertainty on $V$, so that linear 
fits with errors on $V$ only or orthogonal fits with errors on both 
$V$ and $L$ are almost identical. 
Distance errors for the $RL$ relation can be of issue since they affect 
both $R$ and $L$.  However, since $R \propto D$ and $L \propto D^2$, 
distance errors move $RL$ data points along lines of constant surface
brightness.  Since the $RL$ log-slope is ${\sim}0.3$, the effect of 
distance errors is minimised.  This is especially true given the 
larger intrinsic scatter and systematic uncertainties in determining
scale lengths and correcting for internal extinction.

Hubble  types for all  but the  SCII galaxies  were obtained  from the
heterogeneous  NASA  Extragalactic  Data  (NED) base. 
Most SCII
galaxies are not classified in NED and their morphological types were
determined  in a  homogeneous fashion  via a  combination  of eye-ball
examination, B/D  ratio and/or concentration index  (Dale \etal 1999).
Interacting and disturbed galaxies were rejected in all samples.

In addition to the $VRL$ parameters and  Hubble types,  we  have also
considered for our study based on optical images the fully corrected
$\VI$ color, $I$-band central surface brightness, $\mu_{\circ,I}$, and
concentration index.  

For the Shellflow and UMa samples $\VI$ and $\BI$ colors were obtained from
the  original CCD  photometry. For  the  SCII sample $\VI$ colors  were
derived from the SDSS (see Appendix C). For the MAT sample $\BI$ colors
were derived  from RC3  B magnitudes. We converted $\BI$ colors  into $\VI$
colors  using the  relations between  color and  stellar mass-to-light
ratio in  Bell \& de  Jong (2001). These yielded a total of 742
galaxies with available $\VI$.  The central surface brightness
$\mu_{\circ,I}$ are inward extrapolations of the disk fits described 
above and are available for all galaxies in our compilation.

The galaxy light concentration index can be defined as: 
\be
C_{72}=r_{75}/r_{25},
\label{eq:c72}  
\ee where the  radii enclose 75\% and 25\% of the total light
extrapolated to  infinity.  For reference, a pure exponential disk
has $C_{72}=2.8$, regardless of its total  mass or  scale length.
Concentration indices for  the  MAT sample  were  not available.   
In total, 691 galaxies in our compilation have $C_{72}$ estimates.

Altogether, the MAT, SCII, Shellflow, and UMa samples combine for a
total of 1303 separate entries.  The range of physical parameters for
our combined data set is shown in \Fig{histVRL}.  This range is broad
in every respect, yet not complete in any statistical sense.  We have
not endeavored to correct each sample for incompleteness due to the
complex inherent selection biases.  Most parameter distributions 
can be closely approximated by a log-normal function.  The 
relations between morphological type and luminosity, rotation 
velocity, size, color, surface brightness and concentration are 
shown in \Fig{Hubtype}.  The distributions for all the parameters, 
but $\mu_{\circ,I}$, are broadly similar with a gentle decline 
from the highest mean values for the earliest-type systems (Sa)
down to intermediate types (Sbc), settling onto a constant 
mean level for types Sc and later.  This global trend is roughly 
reversed for $\mu_{\circ,I}$. 
Our study of galaxy scaling relations will be extended to 
the $K$-band based on 2MASS data in \se{IRTF} for comparison purposes.  

\subsection{Mean Parameter Relations}{\label{sec:virial}}

We consider the joint projected distribution in each of the planes
defined by a pair of the three log virial variables, $\log \Rd$, $\log
V$, and $\log L_I$ for the four combined samples in
\Fig{ALL_VRLtypes};  the distributions for  the individual  SCII, MAT,
Shellflow, and UMa samples are shown in Appendix A.  Different symbols
identify the full range of spiral Hubble types.

The fits shown as solid black lines in each $VRL$ figure are the
orthogonal linear fits to the full combined data sets of 1303
galaxies. Table 2 gives the results of the orthogonal fits using
propagated errors on both variables 
(see e.g. Akritas \& Bershady 1996) 
for the full sample and for each Hubble type; the errors on the
slopes and zero-points correspond to 1-$\sigma$ deviations assessed by
bootstrap resampling.  The regressions were
performed over the full available range of luminosities, sizes, and
velocities.  The number of galaxies per Hubble type
is shown in parentheses.  The need for more measurements of the
earliest and latest type spirals is obvious. 

We choose orthogonal fits over forward, inverse, or bisector fits for
the following reasons. Forward and inverse fits make the assumption
that scatter exists {\it only} in the dependent variable.  Since it is
not possible for both forward and inverse fits to be correct, bisector
fits, which average the forward and inverse fits, can also not be
correct (indeed, the bisector fit of a perfectly uncorrelated 
distribution of two variables has an absolute slope of 1.)  Furthermore, 
unlike orthogonal fits, forward and inverse fits cannot account 
for measurement uncertainties on both variables simultaneously.

The combined $I$-band data give a $VL$ relation, $V \propto L^\alpha$,
with log-slope $\alpha_I=0.29 \pm 0.01$.  The agreement between the
log-slopes for the $VL$ relations for different samples, as judged from
Figs.~\ref{fig:SCII_VRLtypes}--\ref{fig:UMa_VRLtypes} in Appendix A is
fairly good.  Our $VL$ fit is also a close match to those published by
the original authors (as $M_I \propto \log V$), even
though the minimization techniques can be quite different.  The
Pearson correlation coefficient of the $VL$ relation is $r=0.92$ 
and the conditional distribution of $V|L$ is roughly log-normal.
Our $VL$ relation, which uses optical rotational velocities, is 
a close match to the $I$-band $VL$ relation of Masters \etal (2006) 
for their SFI$^{++}$ catalogue of 807 cluster galaxies when 
allowance is made for line width differences\footnote{The $I$-band
VL relation of Masters \etal (2006) is based on \hi line widths and 
normalised to Sc galaxies.  Application of their transformation from 
\ha to \hi (see their Eq.~2) results in a steeper slope for the $VL$ 
relation based on \hi line widths.  This explains in part the 
difference between the shallower log-slope of our $VL$ relation
(0.29) and theirs (0.32).}.  

The correlations for the $RL$ and size-velocity ($RV$) relations  
are weaker, averaging $r{\sim}0.65$.
The $RL$ relation for the full sample, $R \propto L^\alpha$, has 
a log-slope $\beta_I=0.32 \pm 0.01$. Since the fits of the 
$VL$ and $RL$ relations are more robust than
the $RV$ relation, we adopt an $RV$ relation requiring self-consistency
between the  three relations.   That is, if  $V \propto  L^\alpha$ and
$R\propto  L^\beta$ and $R \propto V^\gamma$, then $\gamma  =
\beta/\alpha$.

Size-dependent distributions are less robust than the $VL$ relation
owing to such factors as, (a) lack of a uniform and
universal definition of scale length (MCH03), (b) a morphological
dependence that we discuss in \se{color}, (c) larger intrinsic scatter
due to the natural dispersion of the spin parameter $\lambda$ (D07),
and (d) susceptibility to surface brightness selection effects.
Whether scale lengths are measured from B/D decompositions or
``marked'' over a specified range of surface brightnesses (while
omitting the bulge region or not) can yield scale differences greater
than 20\% (MCH03).  Again, the overlap between the four samples is too
small to assess systematic errors.  It is therefore difficult to
determine how much of the lower correlation coefficients in the $RL$ and
$RV$ relations is genuine and$/$or due to inadequate scale
lengths\footnote{ The improved reliability of scale length
measurements must come from an extensive program of deep near-infrared
imaging to measure homogeneous and accurate disk scale lengths and
half-light radii for a very large collection of spiral galaxies based
on a single reduction method and high signal-to-noise data.}.  Despite 
potential pathologies with the MAT scale lengths (see \se{data}), the 
scaling relation slopes for this sample and SCII are comfortably close.  
In spite of intrinsic differences between each sample (esp. selection 
functions), we make the assumption that our calibration in
\se{corrections} brought all the samples on the same system.

It has been proposed that the scaling relations may be different for
the fainter, LSB galaxies (e.g., Kauffmann \etal 2003; Shen \etal
2003), where gas fractions are likely to be different. 
Such a departure is not apparent in
the samples that most closely probe the LSB regime (SCII and MAT; see
Figs.~\ref{fig:SCII_VRLtypes} \& \ref{fig:MAT_VRLtypes} in Appendix A).  
Our observations are also corroborated by the study of low mass
dwarf galaxies of Geha \etal (2006) who finds no break in the TF
relations of very faint and brighter spiral galaxies.

In  summary, we  find that  the  mean $I$-band  scaling relations  for
spiral galaxies in Table 2 are:
\be V \prop L_I^{0.29}, \quad \Rx \prop L_I^{0.32}, \quad \Rx
\prop V^{1.10}. 
\label{eq:scaling}
\ee
The uncertainties in the log-slopes for the SCII sample, which we use
as our baseline, are $\pm 0.01$ ($VL$), $\pm 0.02$ ($RL$), and $\pm 0.12$
($RV$) based on 1-$\sigma$ measurement errors.

A study of the volume-corrected $RL$ distribution for 140,000 
late-type SDSS spirals (Shen \etal 2003; hereafter Sh03) has 
yielded log-slopes from $\beta=0.23$ (low $L$) to $\beta=0.53$ 
(high $L$), with scatter ranging from $\sigma_{{\rm ln} R}=0.45$ 
(low $L$) to 0.30 (high $L$).  These results are fully consistent 
with ours; see Table 2 where our $I$-band $RL$ log-slopes agree 
perfectly with those of Sh03. This nearly perfect agreement must
be contrasted in light of various caveats including the fact that 
the size measurement in Sh03 uses the S\'ersic half light 
radius, fit to a circularized SB profile. The median half light 
radius of their sample is 2\arcsec, compared to a mean seeing 
FWHM of 1\farcs5 which is cause for concern.  The morphological
sampling of the SDSS study also differs significantly from that
of our collected samples.  The size measurements are thus 
different from ours and possibly called into question due to 
seeing contamination and poorly matched profile shapes (see e.g. MCH03).

The MAT sample (see \Fig{MAT_VRLtypes}) was also analysed by dJL00 
who attempted to correct for statistical incompleteness; they found 
the range $\beta=0.2-0.3$.  Our (non-corrected) result for the MAT
sample alone is $\beta=0.28 \pm 0.01$ consistent with the upper 
envelope of dJL00.  The completeness correction affects mostly 
the low $L$ distribution.  At a given luminosity, one magnitude 
of surface brightness corresponds to 0.2 dex in $\Rd$, and a 
(complete) population of LSB galaxies would give a shallower 
log-slope in agreement with dJL00.  The effect is subtle, given 
the large intrinsic size errors, but we must keep that caveat 
in mind.

\subsection{The Morphological/Color Dependence of Galaxy Scaling 
 Relations}{\label{sec:color}}

A simple understanding of the morphological dependence of the $VL$ and
RL relations is provided by  \Fig{VRL_IHubfits}.  We detect a slight
and a strong morphological dependence of the $VL$ and $RL$ relations,
respectively.  This is confirmed by examination of Table 2.
The morphological dependence of the $VL$ relation was first discussed 
by Roberts (1978), and revisited by Aaronson \& Mould (1983), Rubin 
\etal (1985), Giraud (1986), Pierce \& Tully (1988), G97, and KFF02 
to name a few.  It goes in the sense of early-type spirals rotating
faster than later ones at most optical luminosities.  
The strong  morphological dependence of the $RL$ relation is such that
earlier-type spirals are more compact than the later ones at most low
to intermediate luminosities.  This trend is reversed for the brighter
galaxies with $L  \geq 10^{11} \lsol$.  For early  type (Sa) galaxies,
the $RL$  log-slope is close  to the nominal  limit of 0.5 for constant
surface  brightness  (since  $\Sigma_\circ \propto  L/\Rd^2$). $RL$
log-slopes are progressively shallower from early- to late-type systems.

We can refer to \Fig{Hubtype} to assess whether a correlation with
morphological type is driven by any other galaxy parameters.  On the 
basis of smallest scatter, the strongest correlations with morphological
type would be with color (see also Roberts \&  Haynes 1994; Fig. 5) 
and velocity. 
Given that color is linked to the mean stellar $M/L$ ratio and star
formation history (SFH) of the galaxy (Bell \&  de Jong  2000; Bell
\etal 2003a; Portinari \etal 2004), the latter might thus be a source
of scatter in the $VL$ relation (e.g. Heavens \&  Jimenez 1999; D07).
One expects the $VL$ scatter in bluer bands to be more sensitive to
contributions of instantaneous star formation activity, while at
(near-)infrared bands the scatter from stellar population effects
mainly reflects the convolved star formation histories.

The analysis of the Nearby Field Galaxy Survey (Jansen \etal 2000) by
KFF02 identified $B-R$ color and \ha equivalent width as the main
drivers of the scatter in the optical $VL$  relation.  Color and \ha
equivalent width are both tributary of star formation histories though
the former depends both on the {\it integrated} and instantaneous star
formation while the latter is a function of the current star formation
rate alone. We do detect a weak color dependence of the
$I$-band $VL$ relation (Figs. \ref{fig:ALL_VRLvmi} \&
\ref{fig:dlogVR_L}).  This result is further confirmed in \se{CR07}
and \Fig{coldvlr}, and independently by P07.  The trends are
such that redder disks rotate faster and are more compact than 
bluer ones.  However, while color may contribute to the $VL$ scatter, 
we shall conclude in \se{CR07} that it is not a dominant source 
of scatter to either the $VL$ or $RL$ relations.

Other than observational errors and scatter in luminosity due to
variations in the SFH, $VL$ scatter is also expected from variations in
the halo concentration  parameter and, to a lesser extent, scatter in
the galaxy spin parameter and disk mass fraction.  A discussion of
the  relative  contributions to  the  $VL$ scatter from SFH and halo
parameters is presented in D07.

According  to  V01,  all  correlations  of $VL$  residuals  with  galaxy
observables detected in blue  bands, including color which affects the
zero-point, vanish in the  infrared.  This important assertion however
rests  on  the  study of  a  small  data  sample and  deserves  closer
attention.  Contrary to findings by  Aaronson \& Mould (1983) and V01,
Rubin \etal (1985) found that the morphological type dependence of the
VL  relation  was  reduced,   but  not  eliminated,  at  infrared  $H$
luminosities.
In the section below, we make similar tests for any variations
of the $VL$ and $RL$ relations at infrared wavelengths.  

\subsection{The Infrared (2MASS) Velocity-Luminosity and Size-Luminosity
Relations}{\label{sec:IRTF}}

The advent of  large-scale infrared surveys such as  2MASS provides us
with  $JHK$  luminosities,  effective  (i.e. half  light)  radii,  and
near-infrared   colors  to   construct   (nearly)  dust-free   scaling
relations.   However, with a  typical surface  brightness limit  of $K
\sim 20$ mag  arcsec$^{-2}$, the 2MASS luminosity profiles  are a full
two magnitudes shallower than  the typical $I$-band surface brightness
profiles  for  our galaxies.   Still,  the  2MASS  data yield  scaling
relations  that are as  tight as  the ones  derived at  $I$-band, with
log-slopes  representative of higher  surface brightness  systems.  So
while 2MASS data may be ill-suited for deep imaging investigations of
galaxies (see e.g. Bell \etal 2003a) and deriving the correct slopes 
of the $VL$ and $RL$ relations (see, however, Kudrya \etal 2003), the 
information that they provide about the scatter of (2MASS) IR scaling 
relations may still be useful.
Because of the 2MASS magnitude limit of 13.5 $K$-mag, only the 360 
brightest SCII galaxies have measured 2MASS magnitudes.  For simplicity,
our analysis of $K$-band scaling relations is restricted to the SCII
sample. Note also that we make use of effective radii, supplied 
in the 2MASS data pipeline, rather than disk scale lengths for this 
sample. For reference, a pure exponential disk has $R_e = 1.678 R_d$.

We correct the IR data in a  manner analogous to the  optical data in
\se{corrections}.   We use  the $K$-band  correction from  Tully \etal
(1998) scaled  using the standard Galactic  extinction curve (Schlegel
\etal 1998) to get $JHK$ extinctions. These corrections remove any
inclination dependences  on the $VL$  and color-$L$ relations.  We
do not find any evidence of an inclination dependence to the values
of $R_{e,K}$ and $\mu_{e,K}$ and thus no corrections to these quantities
were applied.  

Figs.~\ref{fig:2M_types}$-$\ref{fig:2M_SBeff} show the $K$-band $VRL$
relations for the 360 brightest SCII galaxies.  Based on this sample, 
the 2MASS infrared scaling relations are $V \prop L_K^{0.27},  
\ReK \prop L_K^{0.35}$, and  $\ReK \prop  V^{1.29}$. 
Thus the $I$-band and 2MASS $K$-band for the bright SCII sample 
have comparable $VL$ relations, but the log-slopes for the $RL$ 
and  RV relations differ substantially. 
We must keep in  mind that the 2MASS $RL$ and $RV$ log-slopes are biased
against the lower surface brightness SCII galaxies hence the steeper
log-slopes for the $RL$ and $RV$ compared to the $I$-band.  
If we restrict the $I$-band sample of SCII galaxies to the bright 
2MASS sub-sample, we find nearly identical $I$-band log-slopes for
the $VL$ (0.28) and $R_eL$ (0.36) relations.  This confirms that the 
different slopes for the $I$- and $K$-band $RL$ relations result 
purely from selection, rather than a systematic difference between 
the $I$-band disk scale lengths and $K$-band effective radii,  
or putative pathologies with the 2MASS sizes.

Because the $VL$ relation is surface brightness independent, which 
we verify for 2MASS data in \Fig{2M_SBeff}, we can still use these
data to assess the parameter dependence in the $VL$ and, to a lesser 
extent, the (incomplete) $RL$ relation\footnote{Although 2MASS surface
brightness profiles only reach $K \sim 20$ mag arcsec$^{-2}$, the
2MASS $J-K$ color terms are still useful for the separation of
redder and bluer disks since IR galaxy bulge and disk color
gradients are already substantial after one disk scale length
(MacArthur \etal 2004).}.

The $K$-band $VL$ and $RL$ relations show similar dependences on
morphological type as the $I$-band relations (\Fig{VRL_IHubfits}).  
As we see from Figs.~\ref{fig:2M_types}$-$\ref{fig:2MdlogVR_L}
the slope, zero-point and scatter of the $K$-band $VL$ relation  
is fully independent of surface brightness, size, and concentration
parameters.  Any correlation with color, $J-K$ this time, is 
difficult to assess given the paucity of late type systems 
in this sample; the $\VI$ color dependence that was more 
prominent (though barely) in \Fig{dlogVR_L} seems to have 
vanished at NIR wavelengths in \Fig{2MdlogVR_L}.
The $RL$ relation shows the usual dependence on surface brightness 
and no dependence on $J-K$ color.

\subsection{Surface Brightness Independence of the $VL$ 
Relation}{\label{sec:scatter}}

Previous reports of the correlation of $VL$ residuals with disk surface
brightness may  have generated  confusion.  Willick (1999)  reported a
correlation  of  $I$-band $VL$  residuals  with  surface brightness and
compactness for his ``LP10K'' survey of distant cluster galaxies using
a moments fitting method  to determine the exponential disk parameters
even in the  presence of irregularities in the  galaxy light profiles.
In an  attempt to  alleviate subjective fitting  boundaries, Willick's
method  used  the entire  surface  brightness  profile, including  the
bulge, thereby biasing the scale length and central surface brightness
measurements.  Applying  his procedure, we can  reproduce the putative
surface  brightness  dependence  of   the  $VL$  relation  while  proper
bulge-to-disk (B/D) fitting techniques (MCH03) find none.

Prior to this study, the  absence of surface brightness  dependence of
the $VL$ relation had also  been verified at optical bands by Sprayberry
\etal (1995),  Zwaan \etal  (1995), and CR99,  and in the  infrared by
V01. Our Figs.~\ref{fig:ALL_VRLmu},   \ref{fig:dlogVR_L},
\ref{fig:2M_SBeff} and  \ref{fig:2MdlogVR_L} corroborate this evidence
very clearly; namely that there is  no dependence of  the $VL$ relation
scatter on $I$-band central surface  brightness $\mu_{0,I}$ or $K$-band
effective  surface  brightness.

The surface brightness independence of the $VL$ has been interpreted
(e.g. Zwaan \etal 1995) as variation of the dynamical mass-to-light
ratio, $\Upstilde$, with surface brightness for a given total
luminosity.  If the surface brightness decreases, $\Upstilde$
increases in such a way that the $VL$ relation remains independent of
surface brightness.  A dependence of $\Upstilde$ on surface brightness
is indeed expected in CDM-based galaxy structure models; the higher
the disk surface brightness, the higher the contribution of baryons 
to the rotation velocity at, say, 2.2 disk scale lengths (Zavala \etal
2003; P05; D07).  However, reproducing the surface brightness
independence of the $VL$ relation is a non-trivial task, especially 
for the highest surface brightness galaxies (CR99; D07).  Indeed,
for a given disk mass and dark halo profile, a small disk 
(higher surface brightness) would contribute significantly
to $\V22$, the total circular velocity at 2.2 disk scale lengths\footnote{
The maximum velocity measurements collected in our data base correspond
roughly to $\V22$.}.  
In the absence of halo contraction, the contribution of the halo 
to $\V22$, for typical halo concentrations, slighty decreases as
disk size decreases.  The net effect is typically that $\V22$ increases 
as disk size decreases (CR99; D07).  When halo contraction is included 
a smaller disk results in a smaller halo and hence even higher total 
$\V22$ (D07).

Gnedin \etal (2006) circumvent this problem by assuming that at a
given stellar mass, smaller and thus higher surface density disks live
in less massive halos.  This solution is not particularly attractive
as it has neither observational nor theoretical motivation. On the
other hand, the observed dependence of gas-to-stellar mass ratio on
surface brightness is effective in reducing the surface brightness
dependence of the $VL$ relation (FAR00; van den Bosch 2000;
D07). Furthermore, if galaxy formation involves clumpy cold flows,
rather than smooth cooling flows, dynamical friction between the
baryons and dark matter may reverse some (or all) of the expected
effects of halo contraction, thus reducing the surface brightness
dependence of the $VL$ relation (D07).

\section{Residual Correlations of Scaling Relations}{\label{sec:CR07}}

In the spirit of CR99, we now examine below the correlations of the
$VL$, $RL$, and $RV$ residuals from the mean relations.  We define the
residuals for each object $i$ as $\Delta y(i)\equiv y(i)-y_{\rm
fit}(i)$.  The orthogonal fits for the global scaling relations on VRL
were described in \se{virial}. 

Fig.~\ref{fig:3res} shows  the residual correlations  for combinations
of  $\dlogV$, $\dlogR$,  and $\dlogL$  for the  four samples  based on
$I$-band imaging (upper) and SCII sample based on 2MASS $K$-band imaging 
(lower).  
The correlation slopes and their associated
errors are  reported in Table  4 and are  shown at the bottom  of each
panel in  \Fig{3res}. The linear correlation  coefficients, $r$,
are also shown at the top  left corner of each
panel. The  residual correlations reported in Table  4 supersede those
presented in CR99 which differ  slightly due to revised corrections to
the raw observables.

There is close  agreement in the correlated residual  solutions of all
optical samples, with the general solution:
\begin{eqnarray*} 
  \partialVRL & = & -0.07 \pm 0.01 \\ \partialRLV & = &
\phantom{-}0.71 \pm 0.02 \\ \partialVLR & = & \phantom{-}0.31 \pm 0.01.
  \label{eq:CR007}
\end{eqnarray*}
%
While we do  not correct for possible sample  incompleteness, the fact
that  all  of  our   samples,  with  their  very  different  selection
functions,  yield  similar   solutions  suggests  that  any  potential
selection bias  ought to be minimal  or cancel out  (see discussion in
D07).  The  slopes for  the $\dlog V|L-\dlog R|L$ residuals  for all
samples  are   negative  (anti-correlated)  and   low.  This  residual
correlation  is statistically  different  from $\partialvr=0$,  though
only weakly.  Note also that galaxies of all morphological types (and
barredness, not shown here), scatter normally about the zero line (see
Appendix A). 

FAR00 have also computed the VL/RL residuals for the UMa sample 
(\Fig{UMa_VRLtypes}) and found no correlation for both for LSB and 
HSB galaxies, as do we for that sample (see Table 4 and the lower 
left panel of Fig.~\ref{fig:ALLres}). 

The weak surface brightness dependence  of the $VL$ relations  has been
used to infer the ratio of baryonic to dark mass in the inner parts of
galaxies (CR99).  D07 however suggest that changes in the stellar
mass-to-light  ratio, variations  of  baryonic-to-stellar mass  ratio,
bulge formation  prescriptions and  assumptions about the  response of
the dark  matter halo to the  cooling baryons may all  contribute to a
weaker surface brightness dependence of the $VL$ relation.

With $r\simeq0.5$, the $RV$ and $LV$ residuals (middle column of
Fig.~\ref{fig:3res}) are also weakly correlated.
%
%
The $VR$ and $LR$ residual   distributions   (third   column   of
Fig.~\ref{fig:3res})  are  tightly  correlated with  $r\simeq0.9$  (a
little less  at $K$)  and relatively small  slope errors. This, as we
explain below, is simply the $VL$ relation recast in differential form.

The correlations between residuals of the $VRL$ relations give clues to
the contribution of scatter in $V$, $L$, and $R$.  Consider a case where 
scatter exists only in V: then the $\dlog R|V-\dlog L|V$ residuals
will be correlated with a slope equal to the log-slope of the $RL$ 
relation (i.e. ${\sim}0.3$).  Similarly, for scatter only in $L$, 
the slope of the $\dlog V|L-\dlog R|L$ residuals should equal 
the log-slope of the $VR$ relation (i.e. ${\sim}1$); for scatter 
only in $R$, the slope of the $\dlog V|R-\dlog L|R$ residuals should 
equal the log-slope of the $VL$ relation (i.e. ${\sim}0.3$).

Since there is intrinsic scatter in each of the $VL$, $RL$, and $RV$
relations, there must be scatter in at least two of $V$, $L$ and $R$. 
The case of uncorrelated scatter in $R$ and $L$, and no scatter in $V$, 
would result in a positive $\dlog V|L-\dlog R|L$ residual correlation,
uncorrelated $\dlog R|V-\dlog L|V$ residuals, and a weak positive
$\dlog V|R-\dlog L|R$ residual correlation.  This situation is 
in clear conflict with all three observed residual correlations.  
The case with uncorrelated scatter in $V$ and $R$, and no scatter in 
$L$ would result in uncorrelated $\dlog V|L-\dlog R|L$ residuals, 
$\dlog V|R-\dlog L|R$ residuals with slope ${\sim}0.3$, and positively 
correlated $\dlog R|V-\dlog L|V$ residuals.  This latter is indeed
in excellent agreement with the observations.  Thus in the simplest 
interpretation of the residual correlations, the scatter in the $VL$ 
and $RL$ relations is fully independent.  
However, as discussed in D07, reproducing independent $VL$ and $RL$ 
relations is non-trivial as the various theoretically expected 
sources of scatter in $V|L$ and $R|L$ often result in correlated 
$V$ and $R$ scatter.


We conclude this section with a discussion of the color dependence of
residual correlations. As discussed in \se{color}, the $VRL$ relations
show a dependence on color and morphological type.  Since both color
and morphological type depend on luminosity and velocity, a most 
robust way to look for a color dependence of the $VRL$ relations is
to compare the $VRL$ residuals with residuals of the color-luminosity 
and color-velocity relations.  This is shown in
Fig.~\ref{fig:coldvlr}.  The upper panels show the $\VI$ color-luminosity
and $\VI$ color-rotation velocity relations.  The middle panels show the
$VL$ residuals versus color residuals.  The dashed lines show the
predicted correlation slope for scatter in color at a given stellar
mass, assuming $\log M_*/L_I\propto 1.26 (\VI$) from stellar 
population models (Portinari \etal 2004), and that scatter in 
color is uncorrelated with $V$ and $R$.  
For the $\dlog V|L_I-\Delta(\VI)|L_I$ residuals, the observed
correlation is weak but of the same sign as expected (from the 
stellar population models), in agreement with P07.  This further 
confirms that scatter in color is a contributing, but not dominant, 
source of scatter in the $VL$ relation.  However, the 
$\dlog L_I|V-\Delta(\VI)|V$ residuals are prefectly uncorrelated, 
also in agreement with P07.  This is explained if scatter in the 
velocity-stellar mass relation depends on color, in the sense 
that redder galaxies of a given stellar mass rotate slower.  
However, this strikes against theoretical expectations that 
redder galaxies of a given total luminosity should rotate 
faster.

We also find that the $\dlog R|L_I-\Delta(\VI)|L_I$ residuals 
have only a weak correlation, but of opposite sign to that
expected.  This suggests that the scatter in the size-stellar 
mass relation depends on color, with redder galaxies of
a given stellar mass being smaller.  This is expected from 
theoretical galaxy formation models (Bell \etal 2003b). 
Finally the $\dlog R|V-\Delta(V-I)|V$ residuals show no correlation,
but luminosity also does not play any role in this correlation, and
therefore no dependence on colors is expected here. 

\section{Discussion}\label{sec:discussion}

The collection of some of the most extensive existing data bases of galaxy
structural and dynamical parameters at $I$- and $K$-bands has enabled us 
to determine the following major observational results:

\begin{itemize}

\item The $I$-band scaling relations of spiral galaxies, obtained
      by linear regressions of $V$ on $L$ and $R$ on $L$, are
$$ V \prop L_I^{0.29\pm0.01}, \quad \Rd \prop L_I^{0.32 \pm 0.01},
 \quad \Rd \prop V^{1.10}.
$$

\item The 2MASS $K$-band scaling relations of bright SCII galaxies differ 
      slightly (due to selection effects) with
$$ V \prop L_K^{0.27\pm0.01}, \quad \ReK \prop L_K^{0.35 \pm 0.02},
 \quad \ReK \prop V^{1.29}.
$$


The slightly shallower log-slopes of the $VL$ relations from the $I$- 
to $K$-band are expected based on the observed trend of color with
luminosity.  However, the $K$-band $RL$ and $RV$ relations are biased 
to higher slopes due to the surface brightness bias of the 2MASS catalog
for extended objects.  A study of scaling relations based on deeper IR
survey data (e.g. UKIDSS; Hewett \etal 2006) should recover shallower
slopes, comparable to those measured here for $I$-band data, for the 
infrared $RL$ and $RV$ relations.

\item The log-slopes of the $I$- and $K$-band $VL$ and $RL$ relations 
  show a dependence on morphological type, with steeper slopes for 
  earlier type disk galaxies, especially so for the $RL$ relation.

\item  The $I$-band $VL$  relation shows  a dependence on color in the
  direction   expected  from   simple  stellar population  synthesis
  models. However the  $RL$ relation shows a trend  (albeit weak) in the
  opposite  direction, which  is  most simply  interpreted as smaller
  galaxies at a given stellar mass being slightly redder.

\item  The  $VL$  relation  shows  no  dependence on size or surface
  brightness at both $I$ and $K$ bands.

\item The $\dlog V|L-\dlog R|L$ residuals are weakly anti-correlated
  with correlation coefficient $r\simeq-0.2$ and slope
  $-0.07\pm0.01$. Unlike CR99, we showed in Dutton \etal (2007) that 
  this result does not uniquely imply that most spiral disks have 
  sub-maximal fraction of luminous to dark matter.

\item The $\dlog R|V-\dlog L|V$ residuals are weakly correlated with
  correlation coefficient $r\simeq0.5$ and slope $\simeq 0.7$.  
  Together with the weak $\dlog V|L-\dlog R|L$ residual correlation, 
  this suggests that scatter in velocity and size dominate the scatter
  in the $VL$ and $RL$ relations respectively. 

\item The $\dlog V|R-\dlog L|R$ residuals are strongly correlated with
  correlation coefficient $r\simeq 0.9$ and slope $\simeq0.3$.  This
  is the $VL$ relation recast in differential form.

\end{itemize}

The slopes zero points and scatter of the $VL$, $RL$ and $RV$ relations 
can be broadly understood in the context of disk galaxies embedded in dark
matter haloes as discussed in Appendix B and in more detail in Dutton 
\etal (2007).
However, accurately reproducing the zero point and surface brightness
independence of the $VL$ relation, as well as the galaxy number density,
may require a departure from the standard assumptions governing the 
formation of disk galaxies.

In order to make further progress in constraining disk galaxy
formation models from the $VRL$ relations, a large ($N \gta 1000$) 
sample of galaxies with deep multi-wavelength imaging (near-IR is 
essential), high spatial and spectral resolution 1D and 2D 
spectroscopy (with CO, \ha and \hi maps), controlled selection 
criteria, accurate and homogeneous measurements of structural 
parameters, and stellar and baryonic masses, is needed.  Ideally, 
chemical information should be available to constrain 
stellar population effects. 

\vskip -.4cm

\acknowledgments

We acknowledge  useful conversations  and suggestions from  Eric Bell,
Roelof de Jong,  Hans-Walter Rix, Marc Verheijen, and Ben Weiner. 
Michael McDonald and Dan Zucker kindly helped with the extraction 
of Petrosian magnitudes from the SDSS database.
SC wishes to acknowledge his colleagues on the {\it Shellflow} team
(Marc Postman, David Schlegel, and Michael Strauss) for permission to
use and release unpublished data.  SC and LAM acknowledge financial
support from the National Science and Engineering Council of Canada.
AAD acknowledges support from the Swiss National Science Foundation
(SNF).  AD acknowledges support by the US-Israel Bi-National Science
Foundation grant 98-00217, the German-Israel Science Foundation grant
I-629-62.14/1999, and NASA ATP grant NAG5-8218.  DHM acknowledges
support from the National Aeronautics and Space Administration (NASA)
under LTSA Grant NAG5-13102 issued through the Office of Space
Science.  SC would also like to thank the Max-Planck Institut f{\" u}r
Astronomie in Heidelberg and the Max-Planck Institut f{\" u}r
Astrophysik in Munich for their hospitality while part of this paper
was conceived.

This research has made use of (i) the NASA/IPAC Extragalactic
Database (NED) which is operated by the Jet Propulsion Laboratory,
California Institute of Technology, under contract with the National
Aeronautics and Space Administration, as well as NASA's Astrophysics
Data System; (ii), the {\it Two Micron All Sky Survey}, which is a 
joint project of the University of Massachusetts and the Infrared 
Processing and Analysis Center/California Institute of Technology,
funded by the National Aeronautics and Space Administration and
the National Science Foundation; and ($iii$) the {\it Sloan Digital
Sky Survey} (SDSS).  Funding for the creation and distribution of the 
SDSS Archive has been provided by the Alfred P.\ Sloan Foundation, the
Participating Institutions, the National Aeronautics and Space
Administration, the National Science Foundation, the U.S. Department 
of Energy, the Japanese Monbukagakusho, and the Max Planck Society. 
The SDSS Web site is http://www.sdss.org/.
The SDSS is managed by the Astrophysical Research Consortium (ARC)
for the Participating Institutions.  The Participating Institutions 
are the University of Chicago, 
Fermilab, the Institute for Advanced Study, the Japan Participation 
Group, the Johns Hopkins University, Los Alamos National Laboratory,
the Max Planck Institut f\"ur Astronomie (MPIA), the Max Planck 
Institut f\"ur Astrophysik (MPA), New Mexico State University, 
University of Pittsburgh, Princeton University, the United States 
Naval Observatory, and the University of Washington.

\clearpage

\appendix

\section{$I$-band scaling relations and residual correlations for the
  four main samples}{\label{sec:AppendixA}} 

This appendix presents the separate distributions for the $VRL$
scaling relations of the SCII, MAT, Shellflow, and UMa samples.  
These are shown respectively in 
Figs.~\ref{fig:SCII_VRLtypes}-\ref{fig:UMa_VRLtypes}.
The axis limits are the same in all figures and the fits shown 
are those for the full combined sample (see \Fig{ALL_VRLtypes}).

Fig.~\ref{fig:ALLres} shows the residual correlations for combinations
of $\dlogV$, $\dlogR$, and $\dlogL$ for the four samples based on
$I$-band imaging.  The colored point types have the same morphological
dependence as in \Fig{ALL_VRLtypes}.  All four samples give
consistent residual correlations, namely a very weak 
$\dlog V|L-\dlog R|L$ correlation (analagous to the surface 
brightness and size independence of the $VL$ relation), a weak 
positive $\dlog R|V-\dlog L|V$ correlation, and a strong 
positive $\dlog V|R-\dlog L|R$ correlation (the differential 
$VL$ relation). 

\clearpage

\section{On     the     Origin      of     Disk     Galaxy     Scaling
  Relations}\label{sec:origin}
The mean observed trends between  $\Vobs$, $L$ and $\Rd$ at $I$-band
are given by the following $VL$, $RL$, and $RV$ relations:
\be \Vobs \prop L^{0.29}, \quad \Rd \prop L^{0.32}, \quad \Rd \prop
\Vobs^{1.10}.
\label{eq:ave_scaling}
\ee

In the spirit of the  spherical-collapse model (Gunn \& Gott 1972) one
can  define  the virial  radius,  $\Rv$,  of  a collapsed,  virialized
gravitational body as  the radius inside of which  the average density
is  a factor $\Dv$  times the  critical density  of the  Universe (the
value  of $\Dv$  depends on  redshift  and cosmology;  see e.g.,  Eke,
Navarro \& Frenk  1998; Bryan \& Norman 1998). Thus  one can write the
virial mass, $\Mv$, as
\be
\Mv = \frac{4}{3} \pi \Rv^3 \Dv \rhoc,
\label{eq:mrvir}
\ee
where $\rhoc$ is the critical density  of the Universe at a given $z$,
and is given by
\be
\rho_{\rm crit} = {3 \, H(z)^2 \over 8 \, \pi \, G},
\label{eq:rhocrit}
\ee
where $H(z)$ is the Hubble constant at redshift  $z$.  From the virial
theorem, the circular  velocity at the virial radius,  $\Vv$, is given
by
\be
 \Vv^2 = {G \Mv \over \Rv}.
\label{eq:mrv}
\ee
Combining Eqs. \ref{eq:mrvir}-\ref{eq:mrv} and  setting $H(z) =  100 \,h
\kms\Mpc^{-1}$ we obtain
\be \Mv= \Rv^3\, h^{2}\left(\frac{\Dv}{200}\right) \frac{1}{G},
\label{eq:mr}
\ee
\be
\Vv = \Rv\, h \left(\frac{\Dv}{200}\right)^{1/2},
\label{eq:vr}
\ee
and
\be
\Mv = \Vv^3\, h^{-1} \left(\frac{\Dv}{200}\right)^{-1/2} G.
\label{eq:mv}
\ee
Here  $G$ has units  of $(\kms)^2\, \kpc\,\Msun^{-1}$,  $\Mv$ has
units of  $\Msun$, $\Rv$ has units  of $\kpc$, and $\Vv$  has units of
$\kms$.

Thus,  if $\Mv  /  L$, $\Rv  / R_d$,  and  $\Vv /  \Vobs$ are  roughly
constant  for  different galaxy  types  (i.e., not  significantly
correlated  with  either  of  the  virial  quantities),  these  virial
relations are  all that  is required to  explain the  observed scaling
relations of disk galaxies. However,  there are many reasons to expect
significant variation  in $\Mv/L$, $\Rv/\Rd$,  and $\Vv/\Vobs$.  Below
we realize these variations in terms of galaxy formation theory.

The virial mass-to-light ratio can be written as
\be
\label{ratone}
\Mv / L = {\Upsilons \over \fbar \, \egf\,\esf}={\Upsilond \over \md}.
\ee
Here, $\Upsilons \equiv M_*/L$ is defined as the stellar mass-to-light
ratio, $\fbar$ is the baryonic mass fraction of the Universe ($\fbar =
\omb/\Omega_M \simeq  0.15$ for  a $\Lambda$CDM cosmology),  $\egf$ is
the galaxy  formation efficiency that  describes what fraction  of the
baryonic  mass inside the  halo ultimately  ends up  in the  disk, and
$\esf$ is  the star formation efficiency that  describes what fraction
of the disk mass ends up in  the form of stars. We combine $\fbar$ and
$\egf$ into the disk mass  fraction, $\md$, and $\Upsilons$ and $\esf$
into the disk mass to light ratio, $\Upsilond$.

The virial  relation $\Vv  \propto \Mv^{1/3}$ is  thus related  to the
observed $VL$ relation by the quantity
\be  \CVL \equiv  {\Vobs \over  \Vv}\, \left({\Upsilond
  \over \md}\right)^{1/3}. \ee
In order to reproduce the observed $VL$ relation $\Vobs \propto
L^{\alpha_{\rm VL}}$, $\CVL$    must    be    proportional    to
$\Vobs^{\alpha_{\rm  VL} -  1/3}$,  and  has to  reveal  an amount  of
scatter that matches that of the observed $VL$ relation.

In  the  standard model  for  disk formation,  set  forth  by Fall  \&
Efstathiou  (1980), Dalcanton  \etal  (1997) and  MMW98, the  relation
between $\Rd$ and $\Rv$  derives from the acquisition and conservation
of (specific) angular momentum by  the baryons, when cooling to form a
centrifugally  supported  disk.   The  total  angular  momentum  of  a
virialized system is conveniently  expressed by the dimensionless spin
parameter
\be \lambda \equiv {{\Jv} \over \sqrt{2} \Mv \, \Rv \, \Vv} \ee
(Bullock \etal 2001b).  For an exponential disk embedded in a dark
matter halo, the total angular momentum is given by
\be \Jd = 2 \pi \int_{0}^{\Rv} \Sigma(R) V_c(R) R^2 {\rm d}R = \Md \Rd
\Vv f_V 
\ee
with $\Sigma(R)$ and $V_c(R)$ the surface density and circular
velocity of the disk, and
\be f_V = \int_{0}^{\infty} {\rm e}^{-u} \, u^2 \, {V_c(\Rd u) \over
\Vv} \, {\rm d}u 
\ee
(MMW98) a  dimensionless number  (e.g. $f_V=2$ if  $V_c(R)=\Vv$).  The
specific angular momentum  of the disk is related to  that of the halo
via
\be   \ld  =   \lambda   \left(  \frac{\Jd}{\Jv}\right)   \left(
  \frac{\Mv}{\Md}\right)  = \lambda  \left(  \frac{\jd}{\md}\right).
\ee    
If  disk  formation occured  under  conservation  of specific  angular
momentum,  as generally assumed,  then $\ld=\lambda$.   Independent of
this particular assumption one therefore obtains that
\be
\label{eq:rattwo}
{\Rd \over \Rv} = \sqrt{2} \, {\ld \over f_V} \ee
(MMW98). Thus, the virial relation $\Rv \propto \Vv$ is related to the
observed $RV$ relation by the quantity
\be {\cal C}_{\rm RV} \equiv {\ld \over f_V} \, \left( {\Vv \over
\Vobs} \right). 
\ee
In order to reproduce the observed $RV$ relation $\Rd\propto
V^{\alpha_{\rm RV}}$, $\CRV$ must be proportional to
$\Vobs^{\alpha_{\rm RV} - 1}$, and  has to reveal an amount of scatter
that matches that of the observed RV relation.

Similarly, the  virial relation  $\Rv\propto \Mv^{1/3}$ is  related to
the observed $RL$ relation by the quantity
\be  {\cal C}_{\rm  RL} \equiv  {\ld \over  f_V} \,  \left( {\Upsilond
    \over \md} \right)^{1/3}. 
\ee
In order to reproduce the observed $RL$ relation $\Rd\propto
L^{\alpha_{\rm RL}}$,  $\CRL$ must be  proportional to $L^{\alpha_{\rm
    RL} - 1/3}$,  and has to reveal an amount  of scatter that matches
that in the $RL$ relation observed.

Thus, the challenge for galaxy formation theories is to understand how
to meet the  constraints on ${\cal C}_{\rm VL}$, ${\cal C}_{\rm RV}$,
and  ${\cal C}_{\rm RL}$  derived here.  That this is not a trivial
matter becomes clear if one takes into consideration that (i) both
$\md$ and $\Upsilond$ depend extremely sensitively on the efficiencies
of cooling, star formation and feedback, and are thus expected to be
strongly correlated with $\Mv$ and  $\lambda$ (e.g., van den Bosch
2002a), and (ii) that  both  $\Vobs/\Vv$ and $f_V$ depend, in a
convoluted  way, on  $\lambda$, $\md$,  adiabatic contraction, and on
initial  halo concentration (e.g., D07), which itself depends on
halo mass (e.g., Bullock \etal 2001a). 

In summary, while the virial relations of dark matter haloes provide a
natural framework for understanding  the observed scaling relations of
disk galaxies, a direct linkage between the slopes of the observed and
virial relations  ignores the many physical processes  that take place
during galaxy formation.

\clearpage 

\section{An Alternative Derivation of the Tully-Fisher
  relation}{\label{sec:AppendixB}}

Two simple predictions for the Tully-Fisher ($VL$ or TF) relation have been
used in the past: one with a log-slope of $3$ ($-7.5$ in magnitudes),
and the other with a log-slope of $4$ ($-10$ in magnitudes). Here we
compare both predictions and show that they are related to one
another.

From Appendix B, we have
\be
\label{eq:TFthree}
L     =    \Vobs^3     \,h^{-1}    \left(\frac{\Dv}{200}\right)^{-1/2}
\frac{\md}{\Upsilond}   \left(\frac{\Vv}{\Vobs}\right)^{3}  G  \propto
\Vobs^3 \,{\cal C}_{\rm VL}. \ee
This is a TF relation with  log-slope of $3$, if $\CVL$ is independent
of $\Vobs$ (see also Dalcanton  \etal 1997; MMW98; van den Bosch 1998;
Syer, Mao, \& Mo 1999).
%
In contrast,  several authors in the  past have predicted  that the TF
relation should  have a  log-slope of $4$  (e.g., Sargent \etal 1977;
Aaronson, Huchra, \& Mould 1979; Salucci,  Frenk,  \& Persic  1993;
Sprayberry \etal 1995; Zwaan \etal  1995). Their argument goes as 
follows: Assume that the disk surface brightness is described 
by an exponential function of the form
\be
\label{eq:surf_disk}
I(r) = I_0 \exp(-r/\Rd),
\ee
with $\Rd$ the disk scale length and $I_0$ is the disk central
surface brightness. The total luminosity of the disk is
\be
\label{eq:lum_disk}
 L = 2 \pi I_0 \Rd^2.
\ee
Assuming that one measures the rotation velocity $\Vobs$  at a radius
$r = s \Rd$, and that the gas and stars in  the disk move on circular
orbits,
\be
\label{eq:vel_obs}
 \Vobs^2 = {{G M(r)} \over r},
\ee
with $M(r)$ the  total mass within radius $r$.  We can define the total
mass-to-light ratio\footnote{We use the  subscripts ``d'' and ``h'' to
  refer to the disk and dark matter halo, respectively.}
\be
\label{eq:upshat}
\Upstilde(r) \equiv {M(r)\over L} = {\Md(r) + \Mh(r) \over L}, 
\ee
square (\ref{eq:vel_obs}), and then use equations~(\ref{eq:lum_disk}) 
and (\ref{eq:upshat}) to obtain: 
\be
\label{eq:TFfour}
L = \Vobs^4 \,{s^2 \over I_0\,  \Upstilde^2(r)} \,{1 \over 2 \, \pi \,
  G^2} . \ee
This is a TF relation with a log-slope of $4$, as long as
$I_0\,\Upstilde^2(r)$ is  independent of $\Vobs$ and  one measures the
rotation velocity at a  constant number of scale lengths\footnote{Note
  that  this derivation  can be  generalized to  an  arbitrary surface
  brightness  profile  by replacing  the  central surface  brightness,
  $I_0$, with the effective  surface brightness,  $\Ie$, and  the disk
  scale length, $\Rd$, with the effective (i.e. half light) size,
  $\Re$.}.  The  scatter is determined purely by  the variation in
$I_0\,\Upstilde^2(r)$.   This was  emphasized by  Zwaan  \etal (1995),
who, from the finding that both HSB and LSB spirals follow the same TF
relation, concluded that LSB galaxies  must have much larger values of
$\Upstilde^2(r)$ than their HSB counterparts.

The TF relations in Eqs.~(\ref{eq:TFthree}) and~(\ref{eq:TFfour}) thus
predict  a different log-slope.  The cause  of this  apparent paradox
lies in the fact that $I_0$ and $\Vobs$ are {\it  not} independent.
From Eqs.~\ref{eq:mr}, \ref{eq:rattwo} and \ref{eq:lum_disk}, we derive
the relation between surface brightness and observed rotation velocity
\be
\label{eq:IeV}
I_0 = \Vobs\, h\,
\frac{\md}{\Upsilond}\left(\frac{f_V}{\lambda_{\rm{d}}}\right)^2
\left(\frac{\Vv}{\Vobs}\right) \frac{G}{4\pi}. \ee
Substituting this into Eq.~\ref{eq:TFfour} yields a TF relation with a
slope of 3 as long as
\be    \CVL    =   \frac{s^2}{\Upstilde^2(r)}    \frac{\Upsilond}{\md}
\frac{\ld^2}{f_V^2} \frac{\Vobs}{\Vv}  \ee 
is independent of $\Vobs$.  Rather than squaring Eq.~\ref{eq:vel_obs},
a simpler expression results from substituting the relation between
$\Rd$ and $\Vobs$ (see Appendix B) into Eq.~\ref{eq:vel_obs}, thus
resulting in
\be
\label{eq:TFthree_v3}
L     =    \Vobs^3     h^{-1}\frac{s}{\Upstilde(r)}    \frac{\ld}{f_V}
\frac{\Vv}{\Vobs} \left(\frac{\Dv}{200}\right)^{-1/2}.  \ee
This is a TF relation with a slope of 3 as long as
\be  \CVL =  \frac{s}{\Upstilde(r)}  \frac{\ld}{f_V} \frac{\Vv}{\Vobs}
\ee
is independent of $\Vobs$.

Thus the two theoretical  predictions  of  the  TF  relation,  with
log-slopes $3$ and  $4$,  are  directly  related.   However,  while
Eqs.~\ref{eq:TFthree} and \ref{eq:TFthree_v3} have an underlying  
TF slope of 3,  variation of
$\CVL$ with $\Vobs$ (or $L$) will result in a steeper or shallower
slope.  For example, stellar mass-to-light ratios increase with
$\Vobs$ (e.g.  D07), and stellar mass fractions increase with $\Vobs$
(e.g. Kannappan 2004).  This results in the steepest slope for the
stellar mass TF relation, an intermediate slope for the baryonic mass
TF relation, and the shallowest slope for the blue luminosity TF
relation (e.g.  Bell \& de Jong 2001). The slope of the TF relation
will also depend on the velocity measurement, since the shape of
galaxy rotation curves changes systematically with luminosity
(e.g. C97; Catinella \etal 2006).

\bigskip

\section{SDSS Colors for Galaxy 
     Classification}{\label{sec:AppendixC}}

The need for dynamical and color information is crucial for any study 
of galaxy scaling relations.  However, of all the available
samples with {\it homogeneous} dynamical measurements (e.g. rotation 
curves) considered here, digital colors are only available for the 
Shellflow and UMa samples with $VI$ and $BRI$ imaging respectively. 
The MAT and SCII galaxy magnitudes were imaged only at $I$-band.  
In order to augment the available color base, we computed, 
in CR99, a $B-I$ color term for MAT galaxies using B-band 
magnitudes for the RC3 (de Vaucouleurs \etal 1991).  
However, RC3 magnitudes are inherently uncertain with 
$\Delta{m}\simeq0.2$ mag (Courteau 1996).

We can extract SDSS $g$ and $i$ magnitudes  
(as proxy for $V$ and $I$ magnitudes) for the galaxies that 
overlap with our Northern samples.  We do this for Shellflow 
and SCII galaxies in order to map the SDSS system onto the 
Landolt/Cousins 
system upon which Shellflow and SCII magnitudes are based.  
We compare in Figs.~\ref{fig:Shellsloan}  
\& \ref{fig:SCIIsloan} instrumental Petrosian magnitudes from the SDSS  
fifth release against instrumental magnitudes for 31 Shellflow  
and 39 SCII galaxies.  The comparison of instrumental magnitudes avoids
any differences due to extinction and k-corrections.  Besides zero-point 
offsets, the $g$ and $i$ Petrosian magnitudes scale linearly with the 
$V$ and $I$ Cousins magnitudes in Shellflow. The same is true for 
SCII at near-infrared bands.  Because the transformation slopes 
at $I$ band are the same for Shellflow and SCII, we make the 
assumption that the transformation slopes at visual bands will 
also be similar.  We also assume a monochromatic zero-point 
difference between the Shellflow and SCII magnitude systems of 
0.13 mag. 

We compute transformations for the SDSS Petrosian $g$ magnitudes 
into Cousins magnitudes using $g_{\rm Pet} = V_{\rm Land} + 0.36$ 
for the SCII galaxies that overlap with the SDSS/DR5 database. 
The new instrumental ``$V$'' magnitudes from SDSS for 39 SCII 
galaxies were corrected the same way as the $I$-band 
magnitudes but using an extinction dependence $A_V=3.24$, instead  
of $A_I=1.96$, and the Poggianti (1997) formulation for the $V$-band 
k-correction.

\clearpage

\begin{deluxetable}{lcccccc}
\tablecolumns{7}
\tablenum{1}
\tablewidth{0pc}
\tablecaption{Redshift-Distance Galaxy Surveys\label{tab:tab1}}
\tablehead{
\colhead{Sample} & 
\colhead{$N$} & 
\colhead{Gal. type} & 
\colhead{Phot. bands} & 
\colhead{Mag/Diam. limits} & 
\colhead{Redshift limits} & 
\colhead{Rot. measure} \\
\multicolumn{1}{c}{(1)} & \multicolumn{1}{c}{(2)} & \multicolumn{1}{c}{(3)} & \multicolumn{1}{c}{(4)} & 
\multicolumn{1}{c}{(5)} & \multicolumn{1}{c}{(6)} & \multicolumn{1}{c}{(7)} 
}

\startdata
MAT          & 545 & field   & $(B),I$         &  $D_{\rm ESO} \geq$ 1\farcm7 &  $<7000$ \kms   &  \ha \\
SCII         & 468 & cluster & $I$             &  $12 \leq m_{\rm{I}} \leq 17$&  [5000--19,000] &  \ha \\
Shell{\sl flow} & 252 & field& $V,I$           &  $m_B \leq 14.5$             &  [4500,7000]    &  \ha \\
UMa          &  38 & cluster & $B,R,I,K$       &  $m_z \leq 14.5$             &  $20.7 \pm 0.9$ Mpc &  \hi \\
\enddata
\end{deluxetable}

\begin{deluxetable}{cr|ccc|ccc}
\tablecolumns{11}
\tablenum{2}
\tablewidth{0pc}
\tablecaption{Orthogonal fits for the four combined $I$-band samples \label{tab:tab2}}
\tablehead{
\colhead{ } & 
\colhead{ } & 
\colhead{ } & 
\colhead{$\log{V} \ \it{vs} \ \log{L_I}$} & 
\colhead{ } & 
\colhead{ } & 
\colhead{$\log{R_{d,I}} \ \it{vs} \ \log{L_I}$} & 
}
\startdata
Type& N    & slope & \phantom{--}zero-point & $\sigma_{\ln V|L}$ & slope & \phantom{--}zero-point & $\sigma_{\ln R|L}$   \\
ALL & 1303 & $  0.291\pm  0.004$ & $ -0.835\pm  0.039$ &   0.132 &  $  0.321\pm  0.010$ & $ -2.851\pm  0.106$ &   0.325  \\
(V-I) & 742 & $  0.297\pm  0.005$ & $ -0.898\pm  0.056$ &   0.131 &  $  0.306\pm  0.013$ & $ -2.681\pm  0.136$ &   0.304 \\
\hline
Sa  &  117 & $  0.303\pm  0.013$ & $ -0.941\pm  0.140$ &   0.126 &  $  0.550\pm  0.047$ & $ -5.357\pm  0.501$ &   0.375 \\
Sb  &  570 & $  0.288\pm  0.008$ & $ -0.805\pm  0.084$ &   0.132 &  $  0.369\pm  0.020$ & $ -3.373\pm  0.213$ &   0.313 \\
Sc  &  505 & $  0.272\pm  0.006$ & $ -0.648\pm  0.060$ &   0.126 &  $  0.328\pm  0.015$ & $ -2.915\pm  0.156$ &   0.320 \\
Sd  &  111 & $  0.280\pm  0.011$ & $ -0.735\pm  0.111$ &   0.125 &  $  0.254\pm  0.021$ & $ -2.116\pm  0.213$ &   0.289 \\
\enddata
\end{deluxetable}

\begin{deluxetable}{cr|ccc|ccc}
\tablecolumns{8}
\tablenum{3}
\tablewidth{0pc}
\tablecaption{Orthogonal fits for the SCII $K$-band sample \label{tab:tab3}}
\tablehead{
\colhead{ } & 
\colhead{ } & &
\colhead{$\log{V} \ \it{vs} \ \log{L_K}$} & 
\colhead{ } & 
\colhead{ } & 
\colhead{$\log{R_{e,K}} \ \it{vs} \ \log{L_K}$} & 
\colhead{ }
}
\startdata
Type& N    & slope & \phantom{--}zero-point & $\sigma_{\ln V|L}$ & slope & \phantom{--}zero-point & $\sigma_{\ln R|L}$ \\
ALL & 360 & $  0.269\pm  0.007$ & $ -0.692\pm  0.081$ &   0.126 &  $  0.346\pm  0.018$ & $ -3.13\pm  0.19$ &   0.291 \\
\hline
Sa  &  56 & $  0.274\pm  0.021$ & $ -0.721\pm  0.240$ &   0.129 &  $  0.568\pm  0.042$ & $ -5.70\pm  0.47$ &   0.266 \\
Sb  & 131 & $  0.264\pm  0.017$ & $ -0.641\pm  0.185$ &   0.130 &  $  0.428\pm  0.035$ & $ -4.06\pm  0.39$ &   0.265 \\
Sc  & 166 & $  0.255\pm  0.012$ & $ -0.549\pm  0.124$ &   0.118 &  $  0.388\pm  0.025$ & $ -3.54\pm  0.27$ &   0.261 \\
Sd  &   7 & $  0.249\pm  0.039$ & $ -0.504\pm  0.404$ &   0.081 &  $  0.314\pm  0.168$ & $ -2.68\pm  1.80$ &   0.358 \\
\enddata
\end{deluxetable}

\def \partialVRLI {\dlogV{(L_I)} \thinspace / \thinspace\dlogR{(L_I)}}
\def \partialRLVI {\dlogR{(\Vobs)} \thinspace / \thinspace\dlogL_I{(\Vobs)}}
\def \partialRLVK {\dlogR{(\Vobs)} \thinspace / \thinspace\dlogL_K{(\Vobs)}}
\def \partialVLRI {\dlogV{(\Rd)} \thinspace / \thinspace\dlogL_I{(\Rd)}}
\def \partialVRLK {\dlogV{(L_K)} \thinspace / \thinspace\dlogR{(L_K)}}
\def \partialVLRK {\dlogV{(\Rd)} \thinspace / \thinspace\dlogL_K{(\Rd)}}

\begin{deluxetable}{lrccc}
\tablecolumns{5}
\tablenum{4}
\tablewidth{0pc}
\tablecaption{VLR Residual Correlations \label{tab:tab4}}
\tablehead{
\colhead{ } & 
\colhead{ } & 
\colhead{$\Delta\log V|L \ vs \ \Delta\log R|L$} & 
\colhead{$\Delta\log R|V \ vs \ \Delta\log L|V$} &
\colhead{$\Delta\log V|R \ vs \ \Delta\log L|R$} 
}
\startdata
Sample & N     &  slope \hspace{1.0cm} r & slope \hspace{1.0cm} r & slope \hspace{1.0cm} r \\
ALL    & 1303  & $-0.07\pm0.01~(-0.16)$ & $\phantom{-}0.71\pm0.02 ~(+0.53)$ & $\phantom{-}0.31\pm0.01 ~(+0.93)$ \\ 
\hline
MAT    & 545   & $-0.11\pm0.03~(-0.17)$ & $\phantom{-}0.57\pm0.04~(+0.59)$ &$\phantom{-}0.32\pm0.01~(+0.90)$ \\  
SCII   & 468   & $-0.09\pm0.02~(-0.20)$ & $\phantom{-}0.84\pm0.08 ~(+0.53)$ &$\phantom{-}0.32\pm0.01 ~(+0.94)$ \\
Shell{\sl flow}& 252 & $-0.08\pm0.03 ~(-0.18)$ & $\phantom{-}0.78\pm0.09~(+0.52)$ &$\phantom{-}0.32\pm0.01~(+0.93)$  \\
UMa    & 38    & $-0.06\pm0.08~(-0.14)$ & $\phantom{-}0.77\pm0.17~(+0.48)$ &$\phantom{-}0.31\pm0.02 ~(+0.93)$ \\
\hline
SCII K & 360   & $-0.03\pm0.02~(-0.06)$ & $\phantom{-}0.56\pm0.05 ~(+0.52)$ & $\phantom{-}0.28\pm0.01 ~(+0.88)$ \\ 
\enddata
\end{deluxetable} 
\clearpage 

\section{\bf Figure Captions}

{\bf N.B.: The full document, with B\&W and colour figures, is available at: 
  http://www.astro.queensu.ca/$\sim$courteau/papers/VRL2007ApJ.pdf . }

\bigskip 

\begin{figure}
\caption{Distribution of physical parameters for the four combined data sets.
\label{fig:histVRL}}
\end{figure}
\bigskip

\begin{figure}
\caption{Correlation of galaxy observables (disk central surface brightness
 [$\mus  \equiv -2.5\log  I_0$], concentration $C_{72}$, $\VI$ color, 
 luminosity $L_I$, rotation speed $V$, and disk scale length $R_{d,I}$) 
  with morphological type.
  The point types represent each samples: blue squares - Shellflow,
  red stars - SCII, green circles - MAT, and black triangles for UMa.
  There is a weak dependence of all variables on Hubble type, with an
  apparent flattening of the trends for Sc galaxies and later.
  \label{fig:Hubtype}}

\end{figure}
\bigskip

\begin{figure}
\caption{$VRL$ scaling relations for all four samples color-coded by Hubble
  types.  The linear orthogonal fits, shown by the solid lines, are 
  reported in Table 2 for the combined samples; 
  the 2-$\sigma$ observed scatter are given by the dashed lines. 
  \label{fig:ALL_VRLtypes}}
\end{figure}
\bigskip

\begin{figure}
\caption{ 
$VL$ and $RL$ relations color-coded by Hubble types: (left) for the combined
$I$-band sample, (right) for the SCII $K$-band sample (the latter is discussed 
in \se{IRTF}).  The grey dashed lines show the 2-$\sigma$ scatter of the $VL$ 
and  $RL$ relations for the full $I-$ and $K-$band samples.  At $I$-band,
$R=R_d$, the disk scale length; at $K$-band, we used $R=R_e$, the effective radius
of the galaxy supplied by 2MASS. The $VL$ relation shows, on average, 
that earlier-type spirals 
rotate faster than later-types at a given luminosity. The $RL$ relation depends
strongly on morphology with earlier types showing a steeper ($\sim$constant 
surface brightness) slope.
\label{fig:VRL_IHubfits}}
\end{figure}
\bigskip

\begin{figure}
\caption{Same as \Fig{VRL_IHubfits} but with point types representing different
$\VI$ color. The $VL$ relation shows a weak dependence on color.
\label{fig:ALL_VRLvmi}}
\end{figure}
\bigskip

\begin{figure}
\caption{
Same as \Fig{VRL_IHubfits} but with point types representing different
central surface brightness. The $VL$ relation is fully  independent of surface
brightness, whereas the $RL$ relation  shows the trivial  dependence on
surface brightness.
\label{fig:ALL_VRLmu}}
\end{figure}
\bigskip

\begin{figure}
\caption{
Same as \Fig{VRL_IHubfits} but with point types representing different
concentration index.  The $VL$ relation is fully independent 
of concentration, whereas $RL$ relation shows a weak dependence. 
  \label{fig:ALL_VRLci}}
\end{figure}
\bigskip

\begin{figure}
 \caption{
Residuals from the $VL$ (left) and $RL$ (right) relations as a function of
disk  central surface  brightness, disk scale length, rotation velocity,
concentration index, $\VI$ color, and optical morphological type for all 
the galaxies in our sample. Besides the expected dependence of the $RL$ 
relation on $\mu_\circ$, the scatters of the $VL$ and $RL$ relations 
may show weak correlations on color and concentration respectively.    
\label{fig:dlogVR_L}}
\end{figure}
\bigskip

\begin{figure} 
\caption{
$VRL$ scaling relations of the SCII $K$-band sample separated by Hubble
types. The linear orthogonal fits, shown  by the solid lines, are 
reported in Table 3 for the combined samples; the 2-$\sigma$
observed scatter are given by the dashed lines.
\label{fig:2M_types}}
\end{figure} 
\bigskip

\begin{figure}
\caption{
Same as \Fig{2M_types} but with point types representing different
$J-K$ colors. The $VL$ relation  shows a weak dependence on color, while
the $RL$ and RV relations show none.
\label{fig:2M_JmK}}
\end{figure}
\clearpage

\begin{figure}
\caption{
Same as \Fig{2M_types} but with point types representing different
effective  surface brightnesses.   The $VL$ relation is independent
of surface brightness while the $RL$ relation shows a  trivial
dependence on surface brightness.
\label{fig:2M_SBeff}}
\end{figure}
\bigskip

\begin{figure}
\caption{
Residuals from the 2MASS $VL$ (left) and $RL$ (right) relations as a
function of effective  surface  brightness  $\mue$,  effective  size
($\Re$), rotation velocity, concentration index, infrared $J-K$
color, and optical morphological type.  The sample is grossly
incomplete for types later than Sc. 
\label{fig:2MdlogVR_L}}
\end{figure}
\bigskip

\begin{figure}
\caption{
Correlations between the residuals of the $VRL$ relations for the $I$-band
sample  (upper) and $K$-band  sample (lower).   The best  fitting slopes
with 1-$\sigma$ uncertainty and correlation coefficients  are given in
the lower and  upper left corner of each panel.  There  is only a weak
correlation between $V|L$ and  $R|L$ (left) compared to the prediction
for pure disk only model (dashed line with slope $-0.5$).  There is a 
significant correlation between the $L|V$ and $R|V$ residuals (middle), 
and even tighter correlation between  $V|R$   and   $L|R$, which is simply 
the $VL$ relation in differential form.  
\label{fig:3res}}
\end{figure}
\bigskip

\begin{figure}
\caption{
Correlation residuals  of the I-band relations with  $\VI$ color.  The
upper  panels  show  the  color-luminosity (left)  and  color-rotation
velocity (right)  relations. The middle  panels show the  residuals of
the $VL$  relation versus the color  residuals at constant $L$ (left) and
constant $V$ (right).  The lower panels show the  size residuals versus
the color residuals at constant $L$ (left) and constant $V$ (right).  The
dashed lines give predictions based on scatter in  color at  fixed
stellar mass and assuming, based on stellar population models, that
color is related to stellar mass-to-light ratio.
 \label{fig:coldvlr}}
\end{figure}
\bigskip

\begin{figure}
\caption{Scaling relations for SCII galaxies.  Line widths are measured 
 from \ha rotation curves and disk scale lengths are 
 measured using the ``marking the disk'' technique (see text).  
 Luminosities are computed from fully-corrected $I$-band magnitudes. 
 Unlike Dale \etal the luminosities do not include any correction 
 for morphological type dependence.  The solid black lines and reported 
 slopes correspond the orthogonal fits to the combined data set.  The point
 types and colors correspond to the Hubble classification 
 from the authors (Dale \etal 1999). 
 \label{fig:SCII_VRLtypes}}
\end{figure}
\bigskip

\begin{figure}
\caption{Scaling relations for MAT galaxies.  Line widths are measured 
 from \ha rotation curves and disk scale lengths are measured from 
 bulge-to-disk decompositions.  Luminosities are computed from 
 fully-corrected $I$-band magnitudes.  
 The fits are as in \Fig{SCII_VRLtypes}.  Hubble types are taken 
 from the NASA Extragalactic Data Base. 
\label{fig:MAT_VRLtypes}}
\end{figure}
\bigskip

\begin{figure}
\caption{Scaling relations for Shellflow galaxies.  Line widths are 
 measured from \ha rotation curves and disk scale lengths are measured 
 from bulge-to-disk decompositions.  Luminosities are 
 computed from fully-corrected $I$-band magnitudes.  
 Fits and point types are as in \Fig{SCII_VRLtypes}. 
\label{fig:Shell_VRLtypes}}
\end{figure}
\bigskip

\begin{figure}
\caption{
Scaling relations for UMa galaxies.  Line widths are measured from \hi
synthesis rotation  curves and disk  scale lengths are  measured using
the  ``marking  the disk''  technique  (see  text).  Luminosities  are
computed  from fully-corrected  $I$-band magnitudes.   Fits  and point
types are as in \Fig{SCII_VRLtypes}.
\label{fig:UMa_VRLtypes}}
\end{figure}
\bigskip

\begin{figure}
  \caption{Correlation  residuals  from  the  mean relation  at  fixed
    virial  quantity for the  Shellflow, SCII,  MAT, and  UMa samples.
    Symbols are as in \Fig{ALL_VRLtypes}.  Galaxies of all shapes and
    types (barred and un-barred, early  and later types) obey the same
    surface  brightness independent  scaling relations.   The residual
    distribution  for $\partialvr$ of  all three  samples is  flat for
    barred   and  un-barred   galaxies   of  all   Hubble  types   and
    luminosities.  The middle figures  show a weak correlation between
    $L$  and $R$,  and the  figures on  the right  side recast  the VL
    relation in its differential form.
 \label{fig:ALLres}}
\end{figure}
\bigskip

\begin{figure}
\caption{Comparison of raw SDSS (Petrosian) and Shellflow (Cousins) 
 magnitudes. The two photometry data sets are derivable from 
 one another, independent of a color term at least over the  
 Shellflow magnitude range.
\label{fig:Shellsloan}}
\end{figure}
\bigskip

\begin{figure}
\caption{Comparison of raw SDSS (Petrosian) and SCII (Cousins) 
 magnitudes.  The linear transformation of magnitudes holds over 
 nearly 3 magnitudes, independent of a color term.
\label{fig:SCIIsloan}}
\end{figure}

\end{document}